\documentclass[mnsc,nonblindrev]{informs4}
\OneAndAHalfSpacedXI % current default line spacing
\usepackage[left=1in,top=1in,right=1in,bottom=1in]{geometry}
\usepackage{amssymb,amsmath,amsxtra,amstext,amsfonts,xfrac,mathtools}
\usepackage{graphicx,ccaption,booktabs,enumerate,paralist,mdwlist}
\usepackage{dsfont,verbatim,latexsym}
\usepackage[english]{babel}
\usepackage[utf8]{inputenc}
\usepackage[titletoc,toc,page]{appendix}
\usepackage{mathrsfs}
\usepackage{bm}
\usepackage{bbm}
\usepackage{multirow}
\usepackage[table]{xcolor}
\usepackage{longtable,multirow,threeparttable}
\usepackage{float}
\usepackage{tikz,qtree}
\usepackage{ulem}
\usepackage{mwe} 
\usepackage{subfigure}
\usepackage{eurosym}
\usepackage{fancyhdr}
\usepackage{color}
\usepackage[colorlinks=TRUE,citecolor=MyDarkBlue,urlcolor=MyDarkBlue,linkcolor=MyDarkBlue]{hyperref}
\usepackage{natbib}
\usepackage{mathrsfs}
\usepackage{enumitem}
\usepackage{dcolumn}

\usepackage{placeins}

\allowdisplaybreaks
\floatstyle{ruled}
\usepackage{algorithm,algorithmic}
\newfloat{algorithm}{tbp}{loa}
\providecommand{\algorithmname}{Algorithm}
\floatname{algorithm}{\protect\algorithmname}

\definecolor{MyDarkBlue}{RGB}{158,0,0}
\makeatletter
\newcounter{subhyp} 
\let\savedc@hyp\c@hyp

\newcommand{\normhyp}{%
  \let\c@hyp\savedc@hyp % revert to the old one
  \renewcommand\thehyp{\arabic{hyp}}%
} 
\makeatother

\bibpunct[, ]{(}{)}{,}{a}{}{,}%
\def\bibfont{\small}%

\begin{document}

%%%%%%%%%%%%%%%%%%%%%%%%%%%%%%%%%%%%%%%%%%%%%%%%%%%%%%%%%%%%%%%%%%%%%%%%%%%%%%%%%%%%
\RUNTITLE{AI-Generated Metadata for UGC Platforms}
\RUNAUTHOR{Zhang et al.}

\TITLE{The Value of AI-Generated Metadata for UGC Platforms: Evidence from a Large-scale Field Experiment}

%%%%%%%%%%%%%%%%%%%%%%%%%%%%%%%%%%%%%%%%%%%%%%%%%%%%%%%%%%%%%%%%%%%%%%%%%%%%%%%%%%%%

\ARTICLEAUTHORS{%
\AUTHOR{Xinyi Zhang$^1$, Chenshuo Sun$^2$, Renyu Zhang$^3$, Khim-Yong Goh$^4$}
\AFF{
$^1$ National University of Singapore, Singapore, \EMAIL{xinyizhang@u.nus.edu} \\
$^2$ Peking University, Beijing, China, \EMAIL{csun@stern.nyu.edu} \\
$^3$ The Chinese University of Hong Kong, Hong Kong, China, \EMAIL{philipzhang@cuhk.edu.hk}\\
$^4$ National University of Singapore, Singapore, \EMAIL{gohky@comp.nus.edu.sg}
%\\**The order of the first author is based on contribution to this work.
}
}

%These AI-generated titles were placed in the bottom-left corner of the viewer interface and thus rarely noticed by viewers. This design setting allows us to identify the effect of metadata augmentation on content consumption without direct user interaction.

\ABSTRACT{AI-generated content (AIGC), such as advertisement copy, product descriptions, and social media posts, is becoming ubiquitous in business practices. However, the value of AI-generated metadata, such as title, remains unclear on user-generated content (UGC) platforms. To address this gap, we conducted a large-scale field experiment on a leading short-video platform in Asia to provide about 1 million users access to AI-generated titles for their uploaded videos. Our findings show that the provision of AI-generated titles significantly boosted content consumption, increasing valid watches by 1.6\% and watch duration by 0.9\%. When producers adopted these titles, these increases jumped to 7.1\% and 4.1\% respectively. This viewership-boost effect was largely attributed to the use of this generative AI (GAI) tool increasing the likelihood of videos having a title by 41.4\%. The effect was more pronounced for the groups more affected by metadata sparsity. Mechanism analysis revealed that AI-generated metadata improved user-video matching accuracy in the platform’s recommender system. Interestingly, for a video for which the producer would anyway have posted a title, adopting the AI-generated title will decrease its viewership on average, implying that AI-generated titles may be of lower quality than human-generated ones. However, when producers chose to co-create with GAI and significantly revised the AI-generated titles, the videos outperformed their counterparts with either fully AI-generated or human-generated titles, showcasing the benefits of human-AI co-creation. This study highlights the value of AI-generated metadata and human-AI metadata co-creation in enhancing user-content matching and content consumption for UGC platforms.
}
%, thus contributing to the literature on the economic impact of GAI and human-AI collaboration.}

%%%%%%%%%%%%%%%%%%%%%%%%%%%%%%%%%%%%%%%%%%%%%%%%%%%%%%%%%%%%%%%%%%%%%%%%%%%%%%%%%%%%
\maketitle
\section{Introduction}\label{sec:intro}
Generative AI (GAI) and AI-generated content (AIGC) have demonstrated significant values and potentials across industries by efficiently producing high-quality content, such as generating advertising copy, improving customer service, and enhancing media production.\footnote{\url{https://www.mckinsey.com/capabilities/growth-marketing-and-sales/our-insights/how-generative-ai-can-boost-consumer-marketing}.} These GAI tools help streamline content creation, reduce manual effort, and improve output quality. Market research\footnote{\url{https://market.us/report/generative-ai-in-content-creation-market/}.} shows that after adopting GAI, 75\% of marketers reported higher content production, while 79\% observed improvements in content quality. Such an efficiency boost has also fueled the growth of the GAI market, which was valued at USD 11.6 billion in 2023 and is projected to grow to USD 175.3 billion by 2033, with a compound annual growth rate of 31.2\%. 

%AI-generated

On user-generated content (UGC) platforms such as short-video platforms like TikTok, content metadata, such as titles and tags, often include descriptive information about the content. This metadata provides structured details such as themes and subjects that may allow recommender systems to better categorize and understand content \citep{wei2024llmrec}. Research on YouTube \citep{hoiles2017engagement} has shown that improved video metadata correlated with a 25\% increase in video search visibility. A Nielsen report indicates that books with complete metadata related to 2.2 times higher sales than those with incomplete metadata.\footnote{\url{https://pnmais.com/wp-content/uploads/2023/11/Report_NielsenBookData_MVB_Metadata_Frankfurt_2022.pdf}.}

Despite its potential value, metadata is largely sparse for online platforms. \citet{malik2017framework} observe that only 42,000 (i.e., 0.26\%) of 1.6 million YouTube videos had complete metadata. Similarly, a dataset of over 50,000 IMDb movies showed that less than 25\% had complete metadata.\footnote{\url{https://www.kaggle.com/datasets/rajugc/imdb-movies-dataset-based-on-genre?resource=download}.} Other platforms, such as Spotify\footnote{\url{https://soundcharts.com/blog/music-metadata}.} and TikTok,\footnote{A dataset of 1,729 TikTok videos indicated that 45\% of short videos have missing hashtags. See \url{https://www.kaggle.com/datasets/vbradculbertson/tiktok-trending-metadata?select=sug_users_vids1.csv}.} have reported similar challenges. The metadata sparsity issue is primarily driven by the time-consuming nature of metadata generation for content producers. Additionally, because metadata is less visible by users and, hence, does not directly engage them, producers often undervalue its importance \citep{peng2023handling}. Platforms, however, could not mandate metadata generation as it may deter content uploads and eventually harm the platform's ecosystem.\footnote{\url{https://www.multicollab.com/blog/user-generated-content/}.}

The recent emergence of GAI technologies has provided UGC platforms with new solutions to such metadata sparsity challenges. Specifically, leveraging AI-generated content (AIGC) to automatically generate metadata such as titles and hashtags could substantially reduce a content creator's burden to come up with such metadata. This raises an important question of both academic and practical interest: \textit{Does AI-generated metadata deliver value for a UGC platform?} While AI holds promise in generating detailed and relevant metadata \citep{agrawal2023beyond,wei2024llmrec}, the answer to this question is not yet conclusive. As discussed above, metadata is instrumental for user-content matching for UGC platforms, but no causal evidence has been reported in the literature to quantify its downstream effects on content consumption in real-world settings. Furthermore, AI-generated metadata could introduce new challenges, such as hallucination, misinformation, stereotypes, and biased responses \citep{liang2021towards,janus2023}. Hence, AI may generate misleading metadata that misrepresents the content, resulting in poor user-content matching and disengaging the matched users. Therefore, rigorous empirical research is needed to quantify the actual value of AI-generated metadata for UGC platforms. Existing GAI studies \citep{chenzenan2023,sunyi2024,reisenbichler2022frontiers} have largely focused on the impact of AIGC that directly engages users (e.g., online advertising creatives), with the influence of AI-generated metadata mostly overlooked. Similarly, data augmentation studies \citep{wei2024llmrec,ellis2018rfid} have largely overlooked the potential of GAI to generate video metadata.

Another critical question faced by a UGC platform is: \textit{How should platforms best leverage AI-generated metadata to boost content consumption?} Current GAI research \citep{chenzenan2023,art2024} has not provided conclusive evidence on whether AI-generated metadata can fully replace human-generated ones. On one hand, AI may outperform humans by learning from (very) large datasets and identifying patterns overlooked by humans \citep{chenzenan2023}, particularly when humans lack the necessary skills or knowledge for metadata generation. On the other hand, GAI may struggle with unique context-specific cases \citep{longoni2019resistance,granulo2021preference}, where human insights and private information about the content goal and intended audience are crucial for creating context-rich metadata \citep{warehouse2022}. Hence, it remains unclear whether AI-generated metadata, human-generated metadata, or a co-creation of both would best work on UGC platforms. The answer to this question is crucial for a platform to make informed operational decisions on deploying AI-generated metadata in its product.  

\begin{figure}[t]
\centering
\includegraphics[width=0.3\textwidth]{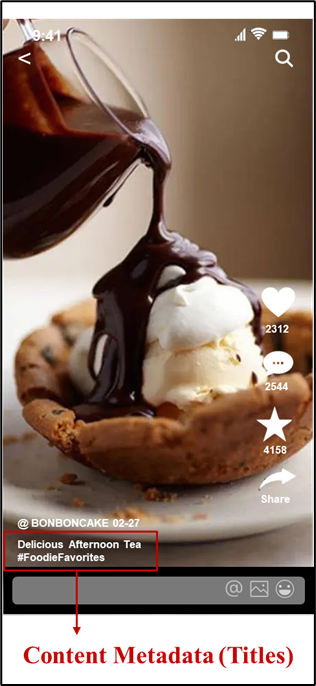}  % Ensure the file name is correct
\caption{Viewer Interface for Watching Videos on Platform A}
\label{fig_viewerinterface}
\end{figure}

To answer these questions, we collaborated with a leading short-video platform in Asia (hereafter “Platform A”) to conduct a large-scale randomized field experiment. Users on Platform A primarily consume videos recommended by the platform. Video titles are the metadata we focus on and rarely noticed by viewers (see Figure \ref{fig_viewerinterface} where video titles were small and positioned at the bottom left corner of the viewers’ interface\footnote{A similar example can be seen on social media platforms like Instagram, where titles and tags are often hidden and require users to click “expand” to view them. Similarly, on e-commerce platforms like Amazon or Taobao, product specifications are often tucked away in dropdown menus, making them less visible unless being actively searched for.}). This metadata is a crucial part of the video profile data used in the recommender system that matches users with relevant videos on Platform A. However, like many UGC platforms, Platform A faces the significant challenge of metadata sparsity. In our dataset (see Section \ref{data_and_variables} for details), only 60.7\% of videos had titles during the pre-treatment period. In response to this challenge, Platform A developed a GAI tool to automatically generate video titles. Specifically, the platform sampled videos with well-matched titles to build a training set, which was then used to fine-tune a multi-modal large-language model (LLM) similar to GPT-4. Once a user uploaded a new video, the GAI tool would capture several frames from the video, extract any text in these frames, and process these combined visual and textual elements as inputs to generate a title. 

In our experiment, users were randomly assigned to either the treatment or the control condition. Content producers in the treatment group received access to AI-generated titles on the video posting page, and could adjust the titles as needed, whereas producers in the control group could only write titles by themselves. Our study covered the treatment period from August 8\textsuperscript{th} to 21\textsuperscript{st}, 2023 and included 2,048,033 producers, each of whom posted at least one video. 

The analyses of our experiment results yield several important insights. We present three findings about \textbf{the effects of AI-generated metadata on content consumption}. First, we find that access to AI-generated video titles significantly boosted video consumption, increasing valid watches\footnote{For videos between 3 and less than 7 seconds, a viewer watch is valid if the watch duration matches the video duration. For videos of 7 seconds or longer, this count is recorded if the watch duration is at least 7 seconds. This metric is designed by the platform to measure the consumption of a video.} by 1.6\%, and watch duration by 0.9\%. When producers adopted these titles, the increase jumped to 7.1\% and 4.1\% respectively. Second, this viewership-boosting effect was likely due to reduced title sparsity on Platform A. We find that access to AI-generated titles increased the likelihood of a video having a title by 41.4\% and tags by 72.4\%. Third, AI-generated titles disproportionately benefited hedonic-content videos (e.g., personal vlogs) and low-skilled content producers more due to their originally sparse video titles. Specifically, utilitarian-content videos (e.g., news and reviews) in the treatment group saw a relative decrease of 3.1\% in valid watches and 3.0\% in watch duration compared to hedonic-content videos. In contrast, low-skilled producers experienced an additional increase of 1.6\% in valid watches, and 1.3\% in watch duration.

%we used the Area Under the ROC Curve (AUC), a metric that measures the system's ability to predict user engagement behaviors. 

Next, we examined the \textbf{mechanism for AI-generated metadata enhancing content consumption}. Prior GAI research \citep[e.g.,][]{chenzenan2023} mostly focuses on prominently displayed AI-generated content that directly engages users. Hence, the mechanisms in these settings cannot be directly applied to explain our results on AI-generated metadata, which are rarely noticed by viewers. Building on studies of data augmentation in recommender systems \citep{wei2024llmrec}, we propose that the enriched metadata helped the platform's recommender system better understand video content and more accurately match videos to fitted users. To validate this hypothesis, we analyzed an additional dataset of 93,618,096 recommendation sessions of the videos produced during our experiment. With this new dataset, we show that the Areas Under the ROC Curve (AUCs) to predict a user's engagement behaviors such as liking, sharing, and following\footnote{While our main analysis focuses on viewership outcomes (e.g., valid watch and watch duration), this additional dataset lacks predicted probabilities for these measures. Instead, it includes predictions for downstream engagement behaviors such as liking, sharing, and following. As these behaviors occur at later stages of the user journey, their accurate predictions imply that viewership outcomes, which occur earlier, are also likely to be predicted accurately.} are significantly higher ($p$ $<$ 0.01) for treatment videos than for control videos. These findings confirm that AI-generated metadata indeed addressed sparse metadata issues, improved user-video matching accuracy, and ultimately drove higher video consumption and engagement.

Finally, we examined \textbf{how content producers should most effectively co-create with AI-generated metadata}. Interestingly, we find that when videos already had human-generated titles, access to AI-generated titles decreased content consumption, with declines of 37.9\% in valid watches and 32.6\% in watch duration. This suggests that AI-generated titles were generally of lower quality than existing human-generated ones. However, when producers chose to co-create with AI to significantly revise AI-generated titles, content consumption would improve. Specifically, each 10\% decrease in textual similarity between AI-generated titles and actual titles adopted by the producers increased the video valid watch by 9.8\% in valid watches and watch duration by 9.2\%. Moreover, lower similarity scores were also associated with richer linguistic attributes, where a 10\% decrease in similarity correlated with a 4.8\% increase in lexical density, 3.7\% in lexical variation, and 1.1\% in entropy. These findings highlight the value of producers and AI co-creating metadata on a UGC platform. Additionally, we surveyed 1,925 treatment group users with open-ended questions on the usage of AI-generated titles. The qualitative feedback pointed to an “inspiration effect,” where AI-generated titles inspired content producers to create better titles, highlighting the potential for human-AI metadata co-creation to boost content consumption.

%the value of content producers and the need to offer them the option to modify AI-generated metadata

% introduces a new approach

In summary, our study leverages AI-generated metadata to improve content-user matching and content consumption on UGC platforms. Our work provides several theoretical and practical contributions. First, we contribute to the research on the economic impact of GAI (\citealt{sunyi2024,art2024}) by exploring the value of a new type of AI-generated content, namely AI-generated metadata. We uncover a new mechanism where AI-generated content enhances engagement by addressing metadata sparsity and improving user-content matching. We also provide new insights on how human-AI metadata co-creation can further improve the accuracy of this matching process. Second, our findings contribute to the growing literature on platform operations \citep{Filippas2023,zeng2023}, shedding light on how to leverage AI-generated metadata to improve user engagement and platform efficiency. Third, we complement the data augmentation literature \citep{peng2023handling,wei2024llmrec} by empirically documenting GAI’s ability to address metadata sparsity and quantifying its economic impacts. Fourth, our study offers unique larege-scale experimental evidence to causally examine the impact of GAI tools on a real-world platform, whereas earlier works are mostly based on observational data or lab-based methods. Lastly, we offer actionable insights for platform managers, emphasizing the importance of using GAI to augment metadata, enhance user-content matching, and ultimately increase content consumption.

The rest of the paper proceeds as follows. Section \ref{sec:literature} reviews the relevant literature. Section \ref{sec:Experiment Design and Data} details our field setting, experimental design, and data. Section \ref{sec:direct_effect} presents the effect of AI-generated metadata on content consumption and the underlying mechanism. Section \ref{sec:human-ai collaboration} explores how content producers collaborate with AI. Section \ref{sec:additional} presents additional analyses and robustness tests. Last, Section \ref{sec:conclu} discusses the practical implications of our research and directions for future research.

\section{Literature Review}\label{sec:literature}
Our paper speaks to three streams of literature: (1) GAI and its collaborations with humans; (2) platform operations; and (3) data augmentation and recommender system.

\subsubsection*{GAI and its Collaborations with Humans.} Our work is most closely connected to research on the economic impact of GAI. Emerging literature has examined its economic impact across various fields including labor market (\citealt{liu2024generatefutureworkai}), firm innovation (\citealt{cheng2022}), marketing (\citealt{chenzenan2023,sunyi2024,reisenbichler2022frontiers}), artwork \citep{art2024}, and knowledge sharing (\citealt{burtch2023}). 

Our contribution to this literature is threefold. First and most importantly, the nascent literature that examines the effect of GAI tools on user engagement has mostly focused on the context where AI-generated content prominently interacts with consumers. For example, the AI-generated summaries (AIGS) (\citealt{sunyi2024}) are displayed to users on product web pages, reducing information search costs and directly affecting consumer behaviors. Similar cases are observed in \citet{chenzenan2023} and 
 \citet{reisenbichler2022frontiers}, which studied advertisement copy that directly impacts consumers’ purchasing decisions. These studies largely attribute the enhanced responses to the enhanced quality of AI-generated content. However, they have overlooked the potential of
GAI to enhance user responses by improving user-content matching through augmented metadata within
recommender systems. Our study fills in this research gap by 
using an innovative experimental design to examine AI-generated metadata, a type of content that is rarely visible to users and does not engage them directly. This design allows us to attribute changes in content consumption primarily to user-content matching rather than user engagement. We thus make a critical contribution by extending the mechanism from enhanced content quality to improved matching accuracy via mitigated metadata sparsity.

Second and method-wise, most prior GAI studies conduct laboratory experiments (\citealt{chenzenan2023}) or treat the launch of ChatGPT or other large language model tools as an event and leverage its timing for econometric identification (usually with Difference-in-Differences method) (e.g., \citealt{art2024,burtch2023}). Our study enhances these studies by using a randomized field experiment to causally assess the impact of GAI tool availability. This experiment provides the exogenous variation in the input of AI-generated metadata in the platform’s recommender system and thus allows us to identify its causal effect on consumption outcomes based on a real-world setting.

Third, our research adds to the growing literature on human-AI collaboration, which has primarily shown that combining human input with AI tools outperforms both full automation and human-only approaches. This has been explored in studies on non-GAI tools (\citealt{anthony2023,boyac2024}) and more recent work on GAI tools (\citealt{chenzenan2023,art2024,wangwen2023}). Research on GAI tools has identified two primary modes of content co-creation: human-revised AI-generated content and AI-revised human-generated content \citep{chenzenan2023}. These studies often compare the linguistic or visual attributes of the content generated by AI, humans, and their co-creation, to understand how these different approaches impact content consumption (\citealt{chenzenan2023,art2024,reisenbichler2022frontiers}). However, because the AI-generated content in these studies is directly shown to users, their findings mainly explain which features enhance user engagement, leaving the open question of what features and how human input can improve user-content matching, which is critical in recommender systems. For example, while emojis can increase user arousal to boost engagement, their complex symbolic nature can limit recommender systems to effectively match content with targeted users. We speak to this open question by designing an experiment to explore AI-generated metadata, a type of AI-generated content that does not directly engage users, to offer a clear understanding of how humans can co-create with AI to enhance user-content matching. Our research advances the understanding of content co-creation process, focusing on human-revised AI-generated content by providing empirical evidence of how human modifications to AI-generated metadata can enhance user-content matching.

\subsubsection*{Platform Operations.} Our research extends the growing literature that address operation problems on online platforms. This literature has examined how to build effective systems for pricing \citep{cui2022ai,zhang2022let},  reviews systems \citep{ruomeng2020reviews}, logistic systems \citep{bai2022value}, social fairness \citep{zhihan2023,clyde2024impact}, content production \citep{zeng2023}, advertisement delivery \citep{ye2023cold}, and content consumption \citep{content2023}. It has also studied how to ensure service quality \citep{cui2020value} and participants’ responses to platform interventions \citep{lysyakov2023threatened,bai2022impacts}. 

We contribute to this stream of research in three ways. First, our work enriches the research that seeks to improve consumers' content consumption on UGC platforms. Prior studies have explored consumer-side interventions such as content prioritization \citep{dukes2024consumption} and personalized content distribution \citep{wei2024llmrec}, as well as producer-side interventions such as financial incentives \citep{kuang2019spillover}, performance feedback \citep{huang2019motivating}, social norms \citep{burtch2018stimulating}, and more recently, AI or GAI tools \citep{he2021effects}. The producer-side interventions largely focus on improving content quality to boost content consumption. Our research takes this
literature a step further by examining the impact of a new operational intervention (i.e., AI-generated metadata) on content consumption outcomes.

Second, our work speaks to the emergent literature that empirically tests the effectiveness of information-based interventions in solving operational problems for online platforms. Examples of prior interventions include providing producers with more information about customers \citep{buell2017creating}, services or products \citep{kesavan2020field,warehouse2022}, and competitors \citep{ruomeng2020reviews,zeng2023}. 
These interventions primarily enhance consumer engagement by improving content production in terms of speed, capacity, and quality. Our study contributes to this literature by introducing a new type of information-based intervention---AI-generated metadata that leverages metadata generation to drive content consumption by enhancing user-content matching. 

\subsubsection*{Data Augmentation and Recommender System.} Recommender system studies have documented that challenges in user-generated video metadata, such as noise, sparsity, and incompleteness, inhibit accurate user-content matching \citep{wei2024llmrec}. To address these issues, data augmentation techniques are developed to enhance data quality. Solutions for handling data sparsity or missing data include, e.g., fuzziness methods \citep{choi2018big}, imputation via supervised learning \citep{ellis2018rfid}, active feature-value acquisition \citep{saar2009active}, and Monte Carlo likelihood estimation \citep{peng2023handling}.

We extend this body of literature in two ways. First, prior research has largely overlooked the potential of GAI to enhance metadata in recommender systems, and even when such attempts exist, they primarily focus on algorithmic aspects without validation in real-world settings \citep[e.g.,][]{agrawal2023beyond,wei2024llmrec}.
Our study contribute by providing a large-scale experimental evidence to quantify the economic value of AI-generated metadata in improving content consumption. Second, unlike past studies that retroactively impute missing data through algorithmic estimation, we propose a proactive approach that enables producers to access AI-generated metadata during metadata generation process. Our findings demonstrate that human-AI co-creation further enhances the value of data augmentation through GAI.

\section{Field Setting, Experiment Design, and Data}\label{sec:Experiment Design and Data}
\vspace{1em}
\subsection{Research Context}\label{sec:Research Context}
We collaborated with one of the largest short-video platforms in Asia (Platform A), which has over 300 million daily active users. Like TikTok, users on the platform can be either content producers or viewers. Producers post short videos on Platform A to enhance viewership/or engagement and attract new followers, aiming to increase advertising opportunities and revenue. Users visit Platform A either to be entertained by videos that catch their interest (organic browse) or to search for specific videos related to a topic (search-oriented browse).

Viewers can consume videos and engage with others for free on Platform A. They engage with producers mainly through viewership, but they can also like videos, leave comments, forward content to others both on and beyond Platform A, and follow producers for long-term video consumption and engagement. The platform generates revenue primarily through online advertising, i.e., disseminating advertising videos to viewers. Therefore, accurately matching the content with viewers to improve video consumption and engagement is crucial to Platform A’s business model.

Content distribution on Platform A is through two primary channels: organic recommendations and search-query-oriented recommendations. Organic recommendations generate a personalized video feed based on a user’s viewing history and preference, catering to those browsing without active searches. Search-query recommendations, on the other hand, respond directly to users'  text-based searches, tailoring content to match users’ specific queries. Here, the video title, which can include hashtags and descriptions, is the only content metadata used by the recommender system to improve content relevance.
\begin{figure}[t]
\centering
\includegraphics[width=0.7 \textwidth]{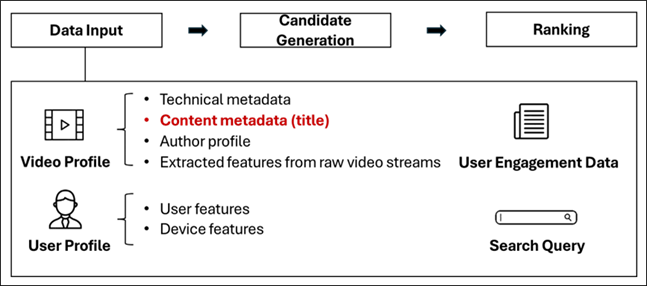}  % Ensure the file name is correct
\caption{Recommender System Workflow on Platform A}
\label{fig_recommendersystem}
\end{figure} 

\begin{figure}[t]
\centering\footnotesize 
\subfigure[Control Group]{
    \begin{minipage}[t]{0.45\textwidth}
        \centering
        \includegraphics[width=\textwidth]{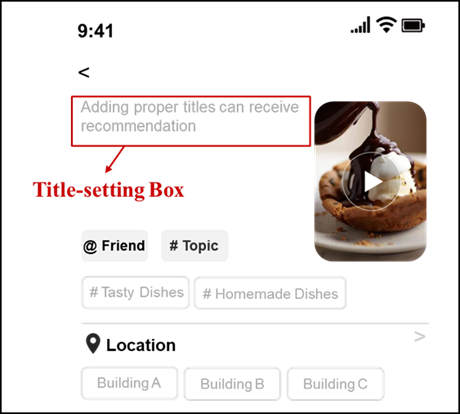} % Ensure equal width
    \end{minipage}%
}%
\hfill
\subfigure[Treatment Group]{
    \begin{minipage}[t]{0.45\textwidth}
        \centering
        \includegraphics[width=\textwidth]{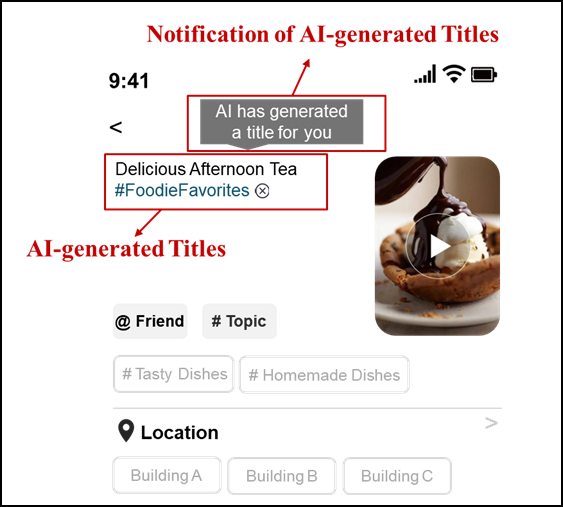} % Ensure equal width
    \end{minipage}%
}%
\centering
\caption[Using AI-generated titles on producers’ video posting page]{\centering Using AI-generated titles on producers’ video posting page}
\label{experiment_design}
\end{figure}

Platform A's recommender system functions in two stages (see Figure \ref{fig_recommendersystem}): candidate generation and ranking \citep{davidson2010youtube}. Both stages rely on four types of data inputs: (1) video profile, (2) user profile, (3) user engagement data (e.g., watch count, duration, likes), and (4) search queries (if any).  Video profile data includes both technical metadata (e.g., file format, upload date, and video duration) and content metadata (i.e., titles), along with the producer profile and extracted features from processing raw video streams (e.g., video category). User profile data include user features (e.g., age and gender) and device features (e.g., device model). In the candidate generation stage, algorithms such as content-based, collaborative filtering, and context-aware methods are used to select relevant video candidates based on user input. A common approach is to pick videos that are closely related to the ones previously watched by the user, utilizing techniques like co-visitation counts and matrix factorization. The ranking stage then takes these candidates and prioritizes them based on the likelihood of user engagement. When posting videos, producers are encouraged to add titles with hashtags or descriptions in the title-setting box (see Figure \ref{experiment_design}(a)). There is no word limit for titles and a producer can also leave it blank, posting a video without a title. Allowing this flexibility helps maintain high upload rates, as mandatory title requirements could complicate the posting process and discourage participation.\footnote{See \url{https://www.multicollab.com/blog/user-generated-content/}.}

Video content metadata (titles) is a crucial part of video profile information in Platform A’s recommender system. Unlike technical metadata (e.g., upload times), which is automatically extracted by Platform A, titles require user input. Although the platform can still recommend videos using other inputs, titles provide more specific context than the broad insights generated by extracted features, such as visual patterns, audio cues, and category classifications.\footnote{Rather than directly processing raw video streams, which is computationally intensive, Platform A uses extracted features from these streams in its recommender system. These features are generated by a separate algorithm within the platform. This design can provide a more efficient video recommendation process.} These features, while useful, often lack the precise contextual information required for accurate recommendations. In contrast, video titles offer structured and concise details about a video’s themes and objects, helping the system better understand content and enhance user-content matching \citep{panniello2016research}. In organic recommendations, for instance, content-based algorithms utilize video titles to recommend videos similar to what users have already watched \citep{adomavicius2008personalization}. For users who frequently watch videos with titles with keywords “cooking” or “recipes,” the system can recommend other videos with related keywords. In search-query recommendations, titles provide direct text matches to user searches, addressing challenges in cross-modality matching by offering clear textual references that improve accuracy. In our study, organic recommendations drove the vast majority of video discovery, and search-oriented recommendations only accounted for less than 1\% of total viewership. Titles are also vital in addressing the cold-start problem,  where new videos without engagement data rely heavily on descriptive titles to be categorized and recommended \citep{wei2024llmrec}. Additionally, titles help mitigate the effects of noisy engagement data, ensuring stable and relevant recommendations even when user engagement is inconsistent due to short video life cycles \citep{davidson2010youtube}.

\begin{figure}[t]
\centering
\includegraphics[width=0.5 \textwidth]{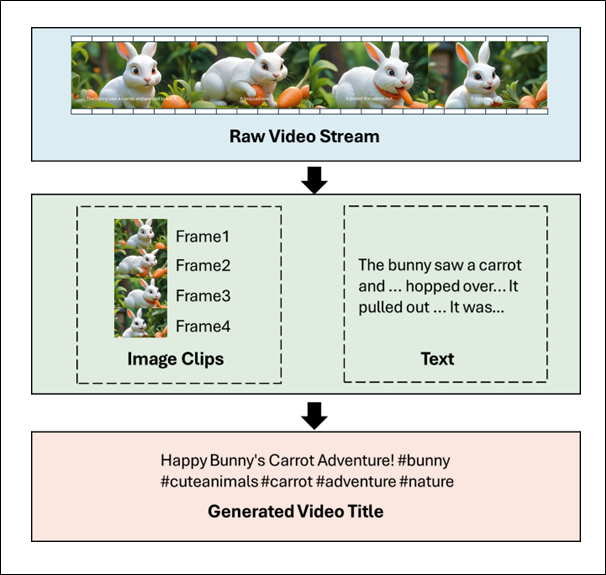}  % Ensure the file name is correct
\caption{Process of Generating Titles}
\label{ai_generated}
\end{figure} 

However, like many UGC platforms, Platform A faces the significant challenge of metadata sparsity. In our dataset (details are presented in Section \ref{data_and_variables}), only 60.7\% of videos had titles during the pre-treatment period. To address this, Platform A introduced GAI tools in July 2023, leveraging GPT-4, a multimodal model capable of processing both text and images. Leveraging transformers with self-attention mechanisms, GPT-4 produces coherent and relevant text. Platform A fine-tuned this model by curating a manually-selected dataset consisting of videos with well-matched titles as the training set. The fine-tuning process allows the model to learn specific patterns in the relationship between video content (both visual and textual) and their corresponding titles on Platform A. To generate metadata for new videos, as shown in Figure \ref{ai_generated}, the platform’s GAI tool captured multiple frames from the video stream, extracting visual elements (e.g., key objects or scenes) and any text present in the frames (e.g., subtitles, on-screen text). This combination of visual and textual data was then fed into the fine-tuned model, which produced a title that best reflects the input content. When new videos were uploaded, the GAI tool captured frames, extracted visual elements and any on-screen text, and fed this data into the model to generate relevant titles automatically. This process allows the platform to generate coherent metadata without requiring human input for every video.

\subsection{Experiment Design}
To causally examine the values of AI-generated metadata, we conducted a field experiment on the video posting page to simulate the metadata input change that feeds into the video recommender system. This experiment lasted from July 20\textsuperscript{th} to August 21\textsuperscript{st}, 2023. Producers involved in our experiment were randomly assigned to the control and treatment groups. Treatment group producers could access an AI-generated title in the title-setting box on the video posting page after they uploaded the video (see  Figure \ref{experiment_design}(b)), and there is a notification next to the AI-generated title indicating that the provided title is generated by AI. In contrast, control group producers could not access such a tool to generate titles via AI (see Figure \ref{experiment_design}(a)). In addition, in 2023, external GAI tools for generating video titles were unlikely to be widely used since major language models did not support video processing, and few video platforms offered such features. Thus, the risk of contamination, where the control group was unintentionally influenced by the experimental intervention, was limited. Treatment producers had the flexibility to either delete, amend, or fully adopt the AI-generated title, and regardless of their choices, viewers could not see any indication in their interface of whether the titles were generated by AI.  In addition, as shown in Figure \ref{fig_viewerinterface}, video titles were small and positioned at the bottom left corner of the viewers’ interface, so we assumed that any observed changes in video consumption and engagement between the treatment and control groups were unlikely to be driven by changes in the visibility of the titles to viewers.

\subsection{Data and Variables} \label{data_and_variables}
Due to some technical issues, Platform A only stored the AI-generated titles between August 8\textsuperscript{th} and August 21\textsuperscript{st} during the experiment. Thus, our dataset was segmented into two periods: (1) pre-treatment period: July 10\textsuperscript{th} to July 19\textsuperscript{th} and (2) treatment period: August 8\textsuperscript{th}
 to August 21\textsuperscript{st}. Our study included $2,048,033$ producers who posted at least one video during our treatment period, with 1,024,940 in the treatment group and 1,023,093 in the control group. During the treatment period, producers in the treatment group uploaded 5,377,560 videos, while those in the control group posted 5,361,424 videos. During the pre-treatment period, only 60.7\% of videos had titles, which indicates that metadata sparsity is prevalent on Platform A. For each producer, we obtained data for video viewership outcomes, producer characteristics, and video characteristics. To accommodate variations in video posting times during the treatment period, we calculated the cumulative viewership outcomes for each video over the first two weeks after its posting \citep{zeng2023}.

Table \ref{tab:variable_definitions} presents the summary of the variables used in our analysis. Our independent variables are the treatment group dummy ($\textit{Treat}_{i}$), coded as 1 if producer $i$ was assigned to the treatment group, and a binary indicator ($\textit{Adopt}_{ij}$) for whether producer $i$ adopted an AI-generated title for video $j$, coded as 1 if the posted video title exactly matched the AI-generated title. Dependent variables, which capture viewers' video consumption, include viewership metrics such as the number of valid watches ($\textit{ValidWatch}_{ij}$), and viewers’ total watch duration in minutes ($\textit{WatchDuration}_{ij}$). To analyze the heterogeneous effects of AI-generated titles, we used two moderator variables: $\textit{Utilitarian}_{ij}$ and $\textit{LowSkill}_{i}$. $\textit{Utilitarian}_{ij}$ indicate whether video $j$ of producer $i$ was a utilitarian-content video, coded as 1 for know-how and news categories\footnote{Know-how categories generally include educational or instructional videos aimed at elucidating practical skills and knowledge. News categories generally include political news and current affairs. The category classifications are developed by Platform A.}, and 0 otherwise. $\textit{LowSkill}_{ij}$ indicate low-skilled producers, coded as 1 if producer $i$'s number of followers was below the median (420 followers)\footnote{We varied this threshold by applying the 30\textsuperscript{th}, 40\textsuperscript{th}, 60\textsuperscript{th}, and 70\textsuperscript{th} percentiles as alternative cutoffs, and the results remained qualitatively consistent.We also employed alternative measurements, such as whether the cumulative number of videos uploaded by producer $i$ exceeds the median (more details are provided in Section \ref{sec:operationalization}).
}, and 0 otherwise. .

To account for various confounding factors, we include a set of control variables ($\textit{Controls}_{ij}$) reflecting producer, video, and day attributes as follows. First, we include producers’ follower counts and a dummy indicating whether they were key opinion leaders to control producer popularity. Second, we include producers’ tenure on the platform (in years), the number of users they followed, and a dummy for multi-homing presence on the focal and rival platforms to control for producers’ experience and/or expertise level. Third, we include gender and provincial location dummies to control for producers’ demographic and geographic variations. Fourth, we include dummy variables indicating whether the video was publicly visible and whether the video was composed of clips or images to account for video type. Fifth, we include the video’s duration and a binary indicator of whether the producer manually sets a video cover to control for video quality. Sixth, we include dummies for video posting dates and categories to control for temporal and categorical variations. Table \ref{tab:summary_statistics} presents the summary statistics of our focal variables. A correlation matrix of these variables is shown in Table \ref{tab:correlation_matrix} of Online Appendix \ref{appendix1}. To protect Platform A's sensitive information,\footnote{The authors have a Non-Disclosure Agreement with Platform A.} the mean values of four key variables (\textit{ValidWatch}, \textit{WatchDuration}, \textit{Follower}, and \textit{Following}) presented in the tables have been scaled by multiplying some positive numbers.

\begin{table}[htbp]\centering \footnotesize  
\caption{Variable Definitions} 
\label{tab:variable_definitions} 
\begin{threeparttable}  
\begin{tabular}{@{\extracolsep{5pt}} l l} 
\\[-1.8ex]\hline 
\hline \\[-1.8ex] 
Variable & Description \\
\hline \\[-1.8ex] 
$\textit{Treat}_{ij}$ & Coded as 1 if producer $i$ of video $j$ was assigned to the treatment group, or else as 0. \\
$\textit{Adopt}_{ij}$ & Coded as 1 if the title of video $j$ posted by producer $i$ exactly matches the AI-generated title. \\
$\textit{ValidWatch}_{ij}$ & Number of valid watches for video $j$ of producer $i$. \\
$\textit{WatchDuration}_{ij}$ & Viewers' total watch duration (in minutes) for video $j$ of producer $i$. \\
$\textit{Utilitarian}_{ij}$ & Coded as 1 if video $j$ of producer $i$ is a utilitarian-content video, or else as 0. \\
$\textit{LowSkill}_{i}$ & Coded as 1 if producer $i$ is a low-skilled producer, or else as 0. \\
$\textit{Follower}_{i}$ & Number of users that follow producer $i$. \\
$\textit{KOL}_{i}$ & Coded as 1 if producer $i$ is a key opinion leader, or else as 0. \\
$\textit{Experience}_{i}$ & Tenure of producer $i$ (in years) on the platform. \\
$\textit{Following}_{i}$ & Number of users that producer $i$ follows. \\
$\textit{Multihome}_{i}$ & Coded as 1 if producer $i$ multihomes on other short video platforms, or else as 0. \\
$\textit{Female}_{i}$ & Coded as 1 if producer $i$ is female, or else as 0. \\
$\textit{Province}_{i}$ & Province location dummies for producer $i$. \\
$\textit{PublicVisible}_{ij}$ & Coded as 1 if video $j$ is publicly visible. \\
$\textit{VideoDuration}_{ij}$ & Duration of video $j$ (in minutes). \\
$\textit{Cover}_{ij}$ & Coded as 1 if producer $i$ manually sets a video cover for video $j$. \\
$\textit{ContentType}_{ij}$ & Coded as 1 if video $j$ is composed of videos (i.e., not images), or else as 0. \\
$\textit{PostDate}_{ij}$ & Video posting date dummies. \\
$\textit{Category}_{ij}$ & First-level category dummies for video $j$ of producer $i$. \\
\hline 
\hline
\end{tabular} 
\begin{tablenotes}\scriptsize
\item \textit{Notes:} All variables are coded as described in the table. Video metadata variables are collected from the platform's system logs. The variables $\textit{WatchDuration}_{ij}$ and $\textit{ValidWatch}_{ij}$ measure engagement metrics.
\end{tablenotes}
\end{threeparttable}  
\end{table}

\begin{table}[htbp]
\centering
\footnotesize
\caption{Summary Statistics of Focal Variables}
\label{tab:summary_statistics}
\begin{threeparttable}
\begin{tabular}{@{\extracolsep{5pt}} lrrrr}
\hline
\hline \\[-1.8ex]
Variable & Mean & SD & Min & Max \\
\hline \\[-1.8ex]
\textit{Treat}           & 0.501 & 0.500 & 0 & 1 \\
\textit{Adopt}           & 0.117 & 0.322 & 0 & 1 \\
\textit{ValidWatch}      & 236.543 & 8,718.901 & 0 & 10,504,166 \\
\textit{WatchDuration}   & 249.240 & 10,995.043 & 0 & 16,021,141.428 \\
\textit{Utilitarian}     & 0.067 & 0.251 & 0 & 1 \\
\textit{LowSkill}        & 0.500 & 0.500 & 0 & 1 \\
\textit{Follower}        & 477.066 & 649.083 & 0 & 5,133 \\
\textit{KOL}             & 0.005 & 0.073 & 0 & 1 \\
\textit{Experience}      & 2.209 & 1.763 & 0.003 & 5.631 \\
\textit{Following}       & 2,940.926 & 50,989.248 & 0 & 37,260,268 \\
\textit{Multihome}       & 0.732 & 0.443 & 0 & 1 \\
\textit{Female}          & 0.649 & 0.477 & 0 & 1 \\
\textit{PublicVisible}   & 0.921 & 0.270 & 0 & 1 \\
\textit{VideoDuration}   & 0.466 & 1.020 & 0 & 43.532 \\
\textit{Cover}           & 0.137 & 0.344 & 0 & 1 \\
\textit{ContentType}     & 0.744 & 0.437 & 0 & 1 \\
\hline
\hline
\end{tabular}
\begin{tablenotes}\scriptsize
\item \textit{Notes}: All variables are calculated based on video-level and producer-level data. SD stands for standard deviation, and Min and Max represent the minimum and maximum values observed for each variable. Values for \textit{ValidWatch}, \textit{WatchDuration}, \textit{Follower}, and \textit{Following} have been scaled.
\end{tablenotes}
\end{threeparttable}
\end{table}

\subsection{Randomization Check}
To verify the randomization effectiveness, we compared treatment producers ($N$=1,024,940) and control producers ($N$=1,023,093) on their pre-treatment video engagement outcomes, producer characteristics, and video attributes. The results of pairwise $t$-tests in Table \ref{tab:randomization_check} show no significant differences between treatment and control groups on these observable attributes. These results confirm that the treatment and control producers in our sample were comparable, suggesting that any difference between conditions after the experiment started should be attributed to our experimental manipulation—that is, whether producers had access to and/or adopted AI-generated titles. 

\begin{table}[htbp]
\centering
\footnotesize
\caption{Randomization Check Results} 
\label{tab:randomization_check} 
\begin{threeparttable}  
\begin{tabular}{@{\extracolsep{5pt}} lrrr} 
\\[-1.8ex]\hline 
\hline \\[-1.8ex] 
Variable & Treatment Producers & Control Producers & $P$-value of $t$-test \\
\hline \\[-1.8ex] 
\textit{ValidWatch}       &  163.055 &  176.388 & 0.248 \\
\textit{WatchDuration}    &  137.386 &  146.698 & 0.453 \\
\textit{Utilitarian}      &    0.065 &    0.066 & 0.238 \\
\textit{LowSkill}         &    0.627 &    0.627 & 0.652 \\
\textit{Follower}         & 1286.531 & 1261.090 & 0.601 \\
\textit{KOL}              &    0.003 &    0.003 & 0.830 \\
\textit{Experience}       &    2.343 &    2.341 & 0.553 \\
\textit{Following}        &  378.537 &  378.720 & 0.817 \\
\textit{Multihome}        &    0.856 &    0.855 & 0.103 \\
\textit{Female}           &    0.610 &    0.610 & 0.597 \\
\textit{PublicVisible}    &    0.940 &    0.941 & 0.259 \\
\textit{VideoDuration}    &    0.465 &    0.465 & 0.897 \\
\textit{Cover}            &    0.155 &    0.156 & 0.135 \\
\textit{ContentType}      &    0.757 &    0.757 & 0.725 \\
\hline 
\hline
\end{tabular} 
\begin{tablenotes}\scriptsize
\item \textit{Notes}: The $p$-value column represents the significance level from a $T$-test comparing the treatment and control groups. Values for \textit{ValidWatch}, \textit{WatchDuration}, \textit{Follower}, and \textit{Following} have been scaled.
\end{tablenotes}
\end{threeparttable}  
\end{table}

\section{Effects of AI-generated Metadata on Content Consumption}\label{sec:direct_effect}
Our investigation began by examining the effects of AI-generated metadata (i.e., titles) on the producers' content consumption outcomes of posted videos. Motivated by past studies (Huang et al. 2021; Sun et al. 2019), we aimed to study two types of causal effects: (1) the effect of treatment (i.e., access to AI-generated titles) on video viewership (intention-to-treat effect, ITT); and (2) the effect of treatment-induced adoptions (i.e., adoption of AI-generated titles) on video viewership (local average treatment effect, LATE). Our unit of analysis was at the producer-video level to capture changes in viewership outcomes for each video uploaded by producers.

\subsection{Effects of Having Access to AI-generated Metadata on Content Consumption}\label{sec:intention_to_treat}
We used the ordinary linear squares (OLS) regression specification with robust standard errors to causally estimate the effects of having access to AI-generated titles on viewership outcomes:
\begin{align}\label{formula:main_ols}
    \textit{Outcome}_{ij} = \beta_0 + \beta_1 \textit{Treat}_{i} + \beta_2 \textit{Controls}_{ij} + e_{ij}
\end{align} where $\textit{Treat}_{i}$ is a binary indicator equal to 1 if the producer $i$ was in the treatment group, $\textit{Controls}_{ij}$ includ all prior-mentioned producer-, video-, and day-level attributes, and $e_{ij}$ is the error term. $\textit{Outcome}_{ij}$ represented our two viewership metrics, include $\textit{ValidWatch}_{ij}$ (the number of valid watches) and $\textit{WatchDuration}_{ij}$ (viewers' total watch duration). All continuous variables in $\textit{Outcome}_{ij}$ were log-transformed, incremented by 1 to account for zero viewership outcomes, following the semi-log approach in \citet{cole2018debt}. Highly-skewed control variables were also log-transformed. 

The model estimation results in Table \ref{tab:ai_generated_titles} show that AI-generated titles boosted content consumption. Specifically, column (1) indicates an increase of 1.6\%\footnote{The marginal effect size is calculated as: Exp(0.016)-1=1.6\%. The same calculation method is applied throughout this paper. }  in valid watches in the treatment group compared to the control group ($\beta_1$ = 0.016, $p$-value $<$ 0.01). Results in column (2) indicate that treatment group videos enhanced watch duration by 0.9\% from the control group ($\beta_1$ = 0.009, $p$-value $<$ 0.01). Given our sample covers 2\% of total platform users, who posted over 10 million videos during our experiment, this result translates to billions of additional valid watches and billions of extra minutes in watch duration across the platform, demonstrating significant economic benefits.

\begin{table}[htbp]\centering \footnotesize  
\caption{Results of Having Access to AI-generated Titles on Content Consumption} 
\label{tab:ai_generated_titles} 
\begin{threeparttable}  
\begin{tabular}{@{\extracolsep{5pt}} lcc} 
\\[-1.8ex]\hline 
\hline \\[-1.8ex] 
Dependent Variable & \textit{ValidWatch} & \textit{WatchDuration} \\
 & (1) & (2) \\
\hline \\[-1.8ex] 
\textit{Treat} & 0.016$^{***}$ & 0.009$^{***}$ \\
 & (0.001) & (0.001) \\
\hline \\[-1.8ex] 
Relative Effect Size & 1.6\% & 0.9\% \\
Controls & YES & YES \\
Observations & 10,738,984 & 10,738,984 \\
R-square & 0.297 & 0.322 \\
\hline 
\hline
\end{tabular} 
\begin{tablenotes}\scriptsize
\item \textit{Notes:} $^{***}$\textit{p}$<$0.01; $^{**}$\textit{p}$<$0.05; $^{*}$\textit{p}$<$0.1. Values in parentheses are robust standard errors.
\end{tablenotes} 
\end{threeparttable}  
\end{table}

To investigate the underlying forces of the boosted viewership outcomes, we examined the textual characteristics of video titles using alternative dependent variables in Equation \eqref{formula:main_ols}. Specifically, we utilized two binary indicators: $Is\_title_{ij}$, which indicates whether video $j$ had a title (either GAI or human-generated), and $Is\_tag_{ij}$, which denotes whether the video title included tags. After re-estimating the OLS Equation \eqref{formula:main_ols} with these variables, the results in Table \ref{tab:video_title_analysis} demonstrate that having access to AI-generated titles increased the likelihood of a video having a title by 41.4\% and enhances the probability of having tags by 72.4\%\footnote{The relative effect size is calculated as 0.244/0.590=0.414 and 0.247/0.341=0.724.}. These results suggest that having access to AI-generated titles effectively reduced metadata sparsity by increasing title and tag completeness, which in turn boosted viewership.

\begin{table}[htbp]\centering \footnotesize  
  \caption{Results of Video Title Characteristics Analysis} 
  \label{tab:video_title_analysis} 
  \begin{threeparttable}  
  \begin{tabular}{@{\extracolsep{5pt}} lcc} 
  \\[-1.8ex]\hline 
  \hline \\[-1.8ex] 
  Dependent Variable & \textit{Is\_title} & \textit{Is\_tag} \\
  & (1) & (2) \\
  \hline \\[-1.8ex] 
  \textit{Treat} & 0.244$^{***}$ & 0.247$^{***}$ \\
  & (0.0003) & (0.0003) \\
  \hline \\[-1.8ex]
  Control Baseline (Mean) & 0.590 & 0.341 \\
  Relative Effect Size & 41.4\% & 72.4\% \\ 
  Controls & YES & YES \\
  Observations & 10,738,984 & 10,738,984 \\
  R-square & 0.165 & 0.126 \\
  \hline 
  \hline
  \end{tabular} 
  \begin{tablenotes}\scriptsize  
    \item \textit{Notes:} $^{***}$\textit{p}$<$0.01; $^{**}$\textit{p}$<$0.05; $^{*}$\textit{p}$<$0.1. Values in parentheses are robust standard errors.
  \end{tablenotes}  
  \end{threeparttable}  
\end{table}

Building on these findings, we next examined whether this viewership-boosting effect was stronger for groups more affected by metadata sparsity. We introduced two moderator variables based on content type and producer skill level, as informed by prior research. According to social exchange theory, digital content creators are driven by motives like personal fulfillment or follower growth for financial gain \citep{waskom2005should}. Utilitarian-content videos (e.g., news and reviews) tend to have more detailed metadata to attract followers, while hedonic-content videos (e.g., personal vlogs) typically lack such detail due to being more focused on self-expression. Additionally, digital divide studies \citep{nattamai2024can} imply that low-skilled producers may undervalue metadata due to a limited understanding of how recommender systems work. Accordingly, we constructed two moderator variables: $Utilitarian_{ij}$, identifying whether a video was utilitarian, and $LowSkill_{i}$, indicating low-skilled producers.

During the pre-treatment period, 61.3\% of utilitarian videos had titles, compared to 57.3\% of hedonic videos. Similarly, 65.3\% of videos from low-skilled producers had titles, compared to 70.5\% of those from high-skilled producers. These statistics align with the theoretical argument above, suggesting that AI-generated metadata may have a more pronounced effect on hedonic content and videos from low-skilled producers. To test this hypothesis, we incorporated these moderator variables and their interaction terms with $Treat_i$ in Equation \eqref{formula:main_ols} to assess their influence.

Results in Table \ref{tab:heterogeneous_effects_video_types} show that utilitarian-content videos in the treatment group experienced a relative decrease in valid watches by 3.1\% ($p$-value$<$0.01), and watch duration by 3.0\% ($p$-value$<$0.01) than hedonic-content videos. Overall, utilitarian-content videos in the treatment group showed a 1.4\%\footnote{This is calculated as: Exp(-0.032+0.018)-1 = -1.4\%.} decrease in valid watches, and a 1.9\% decrease in watch duration, which is likely due to the more detailed human-generated metadata already associated with utilitarian-content videos. In contrast, Table \ref{tab:heterogeneous_effects_producer_types} shows that low-skilled producers, compared to high-skilled ones, experienced an increase of 1.61\% in valid watches ($p$-value$<$0.01), and 1.31\% in watch duration ($p$-value$<$0.01) due to the access to AI-generated titles. These findings align with prior research showing that low-skilled workers disproportionately benefit from GAI tools \citep{chenzenan2023}. Altogether, we find that the viewership-boosting effects of AI-generated metadata are stronger for hedonic-content videos and videos produced by low-skilled producers, due to their originally more sparse metadata.

\begin{table}[htbp]\centering \footnotesize
\caption{Heterogeneous Effects of AI-generated Metadata Access on Content Consumption Across Video Types}
\label{tab:heterogeneous_effects_video_types}
\begin{threeparttable}
\begin{tabular}{@{\extracolsep{5pt}}lcc}
\\[-1.8ex]\hline
\hline \\[-1.8ex]
{Dependent Variable} & {\textit{ValidWatch}} & {\textit{WatchDuration}} \\
 & (1) & (2) \\
\hline \\[-1.8ex]
\textit{Treat} & 0.018$^{***}$ & 0.011$^{***}$ \\
 & (0.001) & (0.001) \\
\textit{Treat} * \textit{Utilitarian} & -0.032$^{***}$ & -0.030$^{***}$ \\
 & (0.004) & (0.004) \\
\hline \\[-1.8ex]
Controls & YES & YES \\
Observations & 10,738,984 & 10,738,984 \\
R-square & 0.297  & 0.322 \\
\hline
\hline
\end{tabular}
\begin{tablenotes}\scriptsize
\item \textit{Notes:} $^{***}$\textit{p}$<$0.01, $^{**}$\textit{p}$<$0.05, $^{*}$\textit{p}$<$0.1. Values in parentheses are robust standard errors. \textit{Utilitarian} is absorbed by the video category dummies in \textit{Controls} and therefore not reported in the table.
\end{tablenotes}
\end{threeparttable}
\end{table}

\begin{table}[htbp]\centering \footnotesize
\caption{Heterogeneous Effects of AI-generated Metadata Access on Content Consumption Across Producer Types}
\label{tab:heterogeneous_effects_producer_types}
\begin{threeparttable}
\begin{tabular}{@{\extracolsep{5pt}}lcc}
\\[-1.8ex]\hline
\hline \\[-1.8ex]
{Dependent Variable} & {\textit{ValidWatch}} & {\textit{WatchDuration}} \\
 & (1) & (2) \\
\hline \\[-1.8ex]
\textit{Treat} & 0.004$^{***}$ & -0.001 \\
 & (0.002) & (0.001) \\
\textit{LowSkill} & -0.783$^{***}$ & -0.755$^{***}$ \\
 & (0.002) & (0.001) \\
 
\textit{Treat} * \textit{LowSkill} & 0.016$^{***}$ & 0.013$^{***}$ \\
 & (0.002) & (0.002) \\
\hline \\[-1.8ex]
Controls & YES & YES \\
Observations & 10,738,984 & 10,738,984 \\
R-square & 0.235 & 0.256 \\
\hline
\hline
\end{tabular}
\begin{tablenotes}\scriptsize
\item \textit{Notes:} $^{***}$\textit{p}$<$0.01, $^{**}$\textit{p}$<$0.05, $^{*}$\textit{p}$<$0.1. Values in parentheses are robust standard errors.
\end{tablenotes}
\end{threeparttable}
\end{table}

\subsection{Results of Adopting AI-generated Metadata on Viewership Outcomes}\label{sec:LATE}
To identify the effect of adopting AI-generated metadata, we cannot simply compare producers who adopted AI-generated titles with those who did not, because omitted variables (e.g., producers’ inherent capability to generate titles) may drive both producers’ decision to adopt AI-generated titles and their subsequent video viewership outcomes. Instead, we used the random assignment of producers to the treatment group ($Treat_i$) as an instrumental variable (IV) for the adoption decision of AI-generated titles \citep{huang2021,sun2019}. We employed the following two-stage least squares (2SLS) regression specification:
\begin{equation}\label{2sls_first}
Adopt_{ij} = \gamma_0 + \gamma_1 Treat_i + \gamma_2 Controls_{ij} + \epsilon_{ij}
\end{equation}
\begin{equation}\label{2sls_second}
Outcome_{ij} = \l_0 + \l_1 \hat{Adopt}_{ij} + \l_3 Controls_{ij} + \eta_{ij}
\end{equation} where $\epsilon_{ij}$ and $\eta_{ij}$ are error terms. In Equation \eqref{2sls_first}, $Adopt_{ij}$ is a binary indicator for whether producer $i$ adopts an AI-generated title for video $j$, and is instrumented with $Treat_i$. In Equation (3), $\hat{Adopt}_{ij}$ refers to the instrumented $Adopt_{ij}$, i.e., the fitted value of $Adopt_{ij}$ from Equation (2). $\beta_1$ is the coefficient of interest indicating LATE. $Treat_i$ is a valid IV for two reasons. First, it satisfies the relevance assumption as only the treatment group can access AI-generated titles, significantly influencing adoption. This is evidenced by a high first-stage $F$-statistic of 1,700,000. Second, it satisfies the exclusion restriction because the treatment assignment is random and should not correlate with other observed or unobserved covariates. Moreover, title generation occurs after video upload and just before posting, removing direct influence on video production.

The main effect results are presented in Table \ref{tab:adopting_ai_titles}. The positive coefficients of $Adopt_{ij}$ indicate that adopting AI-generated titles increased valid watches by 7.1\% ($p$-value$<$0.01), and watch duration by 4.1\% ($p$-value$<$0.01). These findings align with our ITT results but reflect a greater magnitude of effect, demonstrating that adopting AI-generated titles significantly enhances content consumption. Similarly, the coefficients of the interaction term in Table \ref{tab:heterogeneous_effects_video_typesadopt} and Table \ref{tab:heterogeneous_effects_producer_typesadopt} show this effect was more pronounced for hedonic-content videos and low-skilled producers, consistent with our ITT results in Table \ref{tab:heterogeneous_effects_video_types} and Table \ref{tab:heterogeneous_effects_producer_types}. Specifically, utilitarian-content videos showed a relative decrease of 14.0\% in valid watches ($p$-value$<$0.01), and 13.2\% in watch duration ($p$-value$<$0.01) compared with hedonic-content videos. In contrast, low-skilled producers, relative to high-skilled ones, received an incremental increase of 6.7\% in valid watches ($p$-value$<$0.01), and 5.5\% in watch duration ($p$-value$<$0.01) due to the adoption of AI-generated titles. However, the net effect was negative for utilitarian-content videos, with a decrease of 1.2\% in valid watches, and 8.8\% in watch duration. This implies that AI-generated titles, while helpful in addressing video title sparsity issues, may not surpass the quality of some existing human-generated titles.

\begin{table}[htbp]\centering \footnotesize
\caption{Results of Adopting AI-generated Titles on Content Consumption}
\label{tab:adopting_ai_titles}
\begin{threeparttable}
\begin{tabular}{@{\extracolsep{5pt}}lcc}
\\[-1.8ex]\hline
\hline \\[-1.8ex]
Dependent Variable & \textit{ValidWatch} & \textit{WatchDuration} \\
 & (1) & (2) \\
\hline \\[-1.8ex]
\textit{Adopt} & 0.069$^{***}$ & 0.040$^{***}$ \\
 & (0.004) & (0.004) \\
\hline \\[-1.8ex]
Relative Effect Size & 7.1\% & 4.1\% \\
Controls & YES & YES \\
Observations & 10,738,984 & 10,738,984 \\
R-square & 0.297 & 0.322 \\
\hline
\hline
\end{tabular}
\begin{tablenotes}\scriptsize
\item \textit{Notes:} $^{***}$\textit{p}$<$0.01, $^{**}$\textit{p}$<$0.05, $^{*}$\textit{p}$<$0.1. Values in parentheses are robust standard errors.
\end{tablenotes}
\end{threeparttable}
\end{table}

\begin{table}[htbp]\centering \footnotesize
\caption{Heterogeneous Effects of Adopting AI-generated Titles on Content Consumption Across Video Types}
\label{tab:heterogeneous_effects_video_typesadopt}
\begin{threeparttable}
\begin{tabular}{@{\extracolsep{5pt}}lcc}
\\[-1.8ex]\hline 
\hline \\[-1.8ex] 
Dependent Variable & \textit{ValidWatch} & \textit{WatchDuration} \\
 & (1) & (2) \\
\hline \\[-1.8ex] 
\textit{Adopt} & 0.078$^{***}$ & 0.049$^{***}$ \\
 & (0.004) & (0.004) \\
\textit{Adopt} * \textit{Utilitarian} & -0.151$^{***}$ & -0.141$^{***}$ \\
 & (0.019) & (0.018) \\
 \hline \\[-1.8ex]
Controls & YES & YES \\
Observations & 10,738,984 & 10,738,984 \\
R-square & 0.297 & 0.322 \\
\hline
\hline
\end{tabular}
\begin{tablenotes}\scriptsize
\item \textit{Notes:} $^{***}$\textit{p}$<$0.01; $^{**}$\textit{p}$<$0.05; $^{*}$\textit{p}$<$0.1. Values in parentheses are robust standard errors. \textit{Utilitarian} is absorbed by the video category dummies in \textit{Controls} and therefore not reported in the table.
\end{tablenotes}
\end{threeparttable}
\end{table}

\begin{table}[htbp]\centering \footnotesize
\caption{Heterogeneous Effects of Adopting AI-generated Titles on Content Consumption Across Producer Types}
\label{tab:heterogeneous_effects_producer_typesadopt}
\begin{threeparttable}
\begin{tabular}{@{\extracolsep{5pt}}lcc}
\\[-1.8ex]\hline 
\hline \\[-1.8ex] 
Dependent Variable & \textit{ValidWatch} & \textit{WatchDuration} \\
 & (1) & (2) \\
\hline \\[-1.8ex] 
\textit{Adopt} & 0.019$^{***}$ & -0.004 \\
 & (0.007) & (0.006) \\
\textit{Lowskill} & -0.785$^{***}$ & -0.756$^{***}$ \\
 & (0.002) & (0.001) \\
 
\textit{Adopt} * \textit{LowSkill} & 0.065$^{***}$ & 0.054$^{***}$ \\
 & (0.008) & (0.007) \\
 \hline \\[-1.8ex]
Controls & YES & YES \\
Observations & 10,738,984 & 10,738,984 \\
R-square & 0.235 & 0.256 \\
\hline
\hline
\end{tabular}
\begin{tablenotes}\scriptsize
\item \textit{Notes:} $^{***}$\textit{p}$<$0.01; $^{**}$\textit{p}$<$0.05; $^{*}$\textit{p}$<$0.1. Values in parentheses are robust standard errors.
\end{tablenotes}
\end{threeparttable}
\end{table}

\subsection{Mechanism: AI-Generated Metadata Facilitates User-Content Matching}\label{sec:mechanism}
So far, we have shown that AI-generated titles significantly increased viewership outcomes. Next, we explore the mechanism behind this effect. Given the importance of metadata in user-content matching of recommender systems, we hypothesized that AI-generated titles improved video viewership by enhancing user-video matching accuracy. To illustrate, AI-generated titles probably help recommender systems better interpret video content. This enhanced interpretation should enable the system to more accurately predict which users are more likely to engage (e.g., view/like/share videos or follow the producer) and recommend/match the video to these specific users. This improved user-video matching accuracy translates into the higher consumption outcomes observed in Sections \ref{sec:intention_to_treat} and \ref{sec:LATE}. 

To evaluate user-video matching accuracy, we used the Area Under the ROC Curve (AUC), a widely applied metric in recommender system studies \citep{chen2024background,bi2024consumer}. AUC measures how well the model predicts viewer engagement behaviors by comparing the predicted and actual viewer engagement outcomes. A higher AUC indicates more accurate user-video matching. To calculate AUC, we collected a proprietary dataset from the platform’s recommender system, documenting video recommendations from November 1\textsuperscript{st} to November 30\textsuperscript{th}, 2023 for videos posted during our experiment.\footnote{Videos produced during our experiment period were still recommended to users with their titles unchanged after the experiment. Because the platform did not archive the predicted engagement data during our experiment period, we used the data collected in November, 2023.} It included 93,618,096 records with details on predicted engagement probabilities (i.e., like videos, share videos, and follow producers) and actual viewer behaviors for each user-video pair. 

While our main analysis focuses on viewership outcomes (e.g., valid watch and watch duration), this additional dataset does not include predictions for these measures. Instead, it focuses on downstream engagement behaviors, which occur at later stages of the user journey. These predicted behaviors serve as essential inputs for the recommender system to match users with content that aligns with their preferences. Higher AUC values for like, share, and follow indicate that the recommender system effectively predicts user interactions. As these behaviors occur at later stages of the user journey, their accurate predictions imply that viewership outcomes, which occur earlier, are also likely to be predicted accurately. Collectively, a higher AUC for these engagement behaviors reflects improved user-video matching accuracy and supports our hypothesis that AI-generated titles enhance video viewership and engagement through better user-video matching.

To compare whether AUC values differ significantly across treatment and control group videos, we performed 1,000 bootstrap resampling iterations to calculate AUC and $p$-values for both treatment and control videos. The results in Table \ref{tab:auc_comparison} showed significant improvements. The AUC for shares increased from 0.823 in the control group to 0.848 in the treatment group, an improvement of 0.026 ($p$-value$<$0.01). For likes, the AUC rose from 0.892 to 0.921, an increase of 0.029 ($p$-value$<$0.01), and for follows, it increased from 0.867 to 0.887, a rise of 0.019 ($p$-value$<$0.01). These results confirmed that AI-generated titles significantly improved user-video matching accuracy. These findings support our hypothesis that AI-generated titles enhance video viewership and engagement through better user-video matching.

\begin{table}[htbp]\centering\footnotesize
\caption{AUC Comparison}
\label{tab:auc_comparison}
\begin{tabular}{@{\extracolsep{5pt}}lcccc}
\\[-1.8ex]\hline 
\hline \\[-1.8ex] 
{Variable} & {Treatment Group} & {Control Group} & {Difference} & {$P$-value} \\
\hline \\[-1.8ex] 
\textit{Share} & 0.848 & 0.823 & 0.026 & $<$0.01 \\
\textit{Like} & 0.921 & 0.892 & 0.029 & $<$0.01 \\
\textit{Follow} & 0.887 & 0.867 & 0.019 & $<$0.01 \\
\hline
\hline
\end{tabular}
\end{table}

\section{AI vs. Human-Generated Titles}\label{sec:human-ai collaboration}
Beyond the impact of AI-generated metadata on content consumption, we next explored whether and how human content producers can co-create with AI-generated metadata to further enhance consumption outcomes, to address whether content producers should be offered the option to modify AI-generated metadata.

\subsection{Effect of Human-AI Co-creation on Content Consumption}\label{sec:human_ai_viewership}
We began our exploration by comparing the effectiveness of AI-generated titles to human-generated titles. While Section \ref{sec:direct_effect} demonstrated that AI-generated titles boosted video viewership by addressing title sparsity, their impact on videos that already have human-generated titles remained unclear. To investigate this, a direct comparison between titled videos in the treatment and control groups would be misleading. This is because some videos in the treatment group may only be titled because of the access to AI-generated titles, potentially indicating lower producer effort and video quality. Thus, such a direct comparison could underestimate the true effect of accessing AI-generated titles on viewership for titled videos. To address this, we used the propensity score matching (PSM) method with the radius matching algorithm \footnote{We also applied other matching algorithms (e.g., kernel matching) and found robust results.}, employing one-to-many matching to pair each titled video in the treatment group with several of the most “similar” titled videos in the control group based on pre-treatment covariates\footnote{We removed 2,226,922 titled videos (51.56\% of the total titled videos) in our treatment group that did not receive AI-generated titles due to algorithmic issues.} (more details are available in Appendix \ref{appendixpsm}). Using the matched sample, we re-estimated Equation \eqref{formula:main_ols}, applying weights to account for the multiple matches per treated unit. Interestingly, the results in Table \ref{tab:table12} show that videos in the treatment group experienced a decrease of 37.9\%\footnote{The marginal effect is calculated as: Exp(-0.476)-1 = -37.9\%. The same calculation method is applied subsequently.} in valid watches and 32.6\% in watch duration compared to the control group. These results suggest that AI-generated titles may not outperform existing human-generated titles in terms of quality.

\begin{table}[htbp]\centering \footnotesize
\caption{Model Estimation Results for Titled Videos (Matched Sample)}
\label{tab:table12}
\begin{threeparttable}
\begin{tabular}{@{\extracolsep{5pt}}lcc}
\\[-1.8ex]\hline 
\hline \\[-1.8ex] 
Dependent Variable & \textit{ValidWatch} & \textit{WatchDuration} \\
 & (1) & (2) \\
\hline \\[-1.8ex] 
\textit{Treat} & -0.476$^{***}$ & -0.394$^{***}$ \\
 & (0.002) & (0.001) \\
 \hline 
Relative Effect Size & -37.9\% & -32.6\% \\
Controls & YES & YES \\
Observations & 3,885,089 & 3,885,089 \\
R-square & 0.329 & 0.351 \\
\hline 
\hline
\end{tabular} 
\begin{tablenotes}\scriptsize
\item \textit{Notes:} $^{***}$\textit{p}$<$0.01, $^{**}$\textit{p}$<$0.05, $^{*}$\textit{p}$<$0.1. Values in parentheses are robust standard errors. The control baseline used here is calculated based on the matched sample.
\end{tablenotes}
\end{threeparttable}
\end{table}

To further explore this phenomenon, we next analyzed the heterogeneous effect of textual similarity on viewership. We followed the text-mining literature \citep{burtch2022peer} and employed a well-established approach to construct textual similarity. Specifically, we employed the cosine distance between numeric representations of textual content in vector space. This measure is constructed using term-frequency inverse-document frequency (TF-IDF\footnote{TF-IDF is a scaled matrix that captures how frequently each term appears in a document relative to its frequency across all documents in the dataset. This scaling down-weights common words that are less informative for distinguishing between documents, thus emphasizing words that are unique to specific documents. If a word is highly unique and only appears in a single document, its impact is preserved, while words common across many documents are given less weight. Calculating cosine distances between document vectors in this TF-IDF derived space effectively measures similarity in terms of distinctive word usage.}) within an embedding space derived from a broader corpus of video titles in both the treatment and control groups in our sample. Using the TF-IDF algorithm, we computed cosine similarities between AI-generated titles and actual video titles adopted by producers ($Similarity_{ij}$). We then incorporated this similarity measure ($Similarity_{ij}$) into Equation \eqref{formula:main_ols} by adding an interaction term between $Treat_i$ and $Similarity_{ij}$:\footnote{The main effect of $ \textit{Similarity}_{ij} $ is not included in this specification because it would be absorbed by the interaction term $ (\textit{Treat}_i \times \textit{Similarity}_{ij}) $ due to collinearity.}
\begin{align}\label{formula:similarity_ols}
    \textit{Outcome}_{ij} = \alpha_0 + \alpha_1 \textit{Treat}_{i} + \alpha_2  (\textit{Treat}_i \times \textit{Similarity}_{ij}) + \alpha_3 \textit{Controls}_{ij} + \mu_{ij} 
\end{align}

The negative coefficient of the interaction term in Table \ref{tab:heterogeneous_similarity_effects} indicates that a 10\% increase in similarity between AI- and human-generated titles led to an 9.8\% decrease in valid watches\footnote{The marginal effect size is calculated as: Exp((-1.026)*0.1)-1 = -9.8\%.}, and a 8.2\% decrease in watch duration. However, interestingly, when considering the positive coefficient for $Treat_i$, we find that treatment group videos with less than 20.8\%\footnote{This is calculated as 0.213/1.026 = 20.8\%. The same calculation method is applied subsequently. } and 20.9\% similarity (i.e., low overlap with AI-generated titles) outperformed the control group in both valid watches and watch duration. These findings suggest that AI-generated titles may be of lower quality than human-generated titles, and thus higher similarity to these titles tended to reduce viewership. However, when content producers revised AI-generated titles—resulting in lower similarity—the negative effect diminished, and treatment group videos ultimately performed better than those in the control group. These findings suggest that while AI-generated titles reduced production costs and addressed title sparsity, producers should revise these titles rather than adopt them without changes. This aligns with prior research showing that human-AI collaboration outperformed both full automation and human-only approaches \citep{boyac2024}, emphasizing the benefits of combining human judgment with AI efficiency \citep{chenzenan2023,art2024}.

\begin{table}[htbp]\centering \footnotesize
\caption{Heterogeneous Estimation Results for Titled Videos (Matched Sample)}
\label{tab:heterogeneous_similarity_effects}
\begin{threeparttable}
\begin{tabular}{@{\extracolsep{5pt}}lcc}
\\[-1.8ex]\hline 
\hline \\[-1.8ex] 
Dependent Variable & \textit{ValidWatch} & \textit{WatchDuration} \\
 & (1) & (2) \\
\hline \\[-1.8ex] 
\textit{Treat} & 0.213$^{***}$ & 0.178$^{***}$ \\
 & (0.002) & (0.002) \\
\textit{Treat} * \textit{Similarity} & -1.026$^{***}$ & -0.852$^{***}$ \\
 & (0.003) & (0.002) \\
\hline \\[-1.8ex]
Controls & YES & YES \\
Observations & 3,885,089 & 3,885,089 \\
R-square & 0.382 & 0.393 \\
\hline 
\hline
\end{tabular} 
\begin{tablenotes}\scriptsize
\item \textit{Notes:} $^{***}$\textit{p}$<$0.01, $^{**}$\textit{p}$<$0.05, $^{*}$\textit{p}$<$0.1. Values in parentheses are robust standard errors.
\end{tablenotes}
\end{threeparttable}
\end{table}

There is an endogenity concern that the videos with significantly revised AI-generated titles, resulting in lower similarity between the AI-generated titles and actual titles adopted by producer, may inherently indicate higher producer effort and video quality. In other words, the observed outcomes could stem from these underlying differences rather than the benefits of human-AI co-creation. To address this issue, we employed the propensity score matching (PSM) method with a radius matching algorithm, utilizing one-to-many matching.  Specifically, for each titled video in the treatment group with cosine similarity to AI-generated titles below 20\%, we matched it with several control group videos that were most similar based on pre-treatment covariates (details are presented in Online Appendix \ref{appendixpsm2}). Using the matched sample, we re-estimated Equation \eqref{formula:main_ols} and applied weights to account for multiple matches per treated unit. The results in Table \ref{tab:table_cosine_match} show that videos in the treatment group outperformed those in the control group in terms of valid watches and watch duration. These findings suggest that the observed higher viewership outcomes are likely driven by human-AI co-creation.

\begin{table}[htbp]\centering \footnotesize
\caption{Model Estimation Results for Titled Videos with Lower Textual Similarity (Matched Sample)}
\label{tab:table_cosine_match}
\begin{threeparttable}
\begin{tabular}{@{\extracolsep{5pt}}lcc}
\\[-1.8ex]\hline 
\hline \\[-1.8ex] 
Dependent Variable & \textit{ValidWatch} & \textit{WatchDuration} \\
 & (1) & (2) \\
\hline \\[-1.8ex] 
\textit{Treat} & 0.153$^{***}$ & 0.121$^{***}$ \\
 & (0.002) & (0.002) \\
 \hline 
Relative Effect Size & 16.5\% & 12.9\% \\
Controls & YES & YES \\
Observations & 3,533,720 & 3,533,720 \\
R-square & 0.242 & 0.269 \\
\hline 
\hline
\end{tabular} 
\begin{tablenotes}\scriptsize
\item \textit{Notes:} $^{***}$\textit{p}$<$0.01, $^{**}$\textit{p}$<$0.05, $^{*}$\textit{p}$<$0.1. Values in parentheses are robust standard errors. The control baseline used here is calculated based on the matched sample.
\end{tablenotes}
\end{threeparttable}
\end{table}

\subsection{Effect of Human-AI Co-creation on Lexical Richness}\label{sec:human_ai_linguistic}

To better understand how increased human input boosted video viewership, we used lexical richness, a key linguistic concept in language studies, signaling information quality, as our alternative dependent variable. We followed prior research \citep{qiao2020financial} and measured it through multiple dimensions, including lexical density ($LexicalDensity_{ij}$), lexical variation ($LexicalVariation_{ij}$), and entropy ($Entropy_{ij}$). Lexical richness is an important proxy for cognitive effort in text crafting and a signal of information quality. For example, \citet{goes2014popularity} used lexical density to measure the informational value of reviews. Lexical density is the proportion of content words (such as nouns, verbs, and adjectives) to the total number of words in a text. Lexical variation is the ratio of unique words to the total number of words. Entropy quantifies text unpredictability, computed as:
\begin{equation}
Entropy = -\sum_{k=1}^n P_k \log P_k
\label{eq:entropy}
\end{equation}
where $P_k$ is the probability of each unique word. We also include sentence length as an additional confounder in this analysis. The results in Table \ref{tab:lexical_richness_results} show that each 10\% increase in similarity score related to a 4.8\%,\footnote{The marginal effect size is calculated as: (-0.245*0.1)/0.509 = -4.8\%.} 3.7\%, and 1.1\% decrease in lexical density, lexical variation, and entropy, respectively. These results show that titles with lower similarity tend to be more descriptive and information-rich. Such titles may provide clearer context and relevant keywords that align more effectively with the video content, improving the recommender system’s ability to match videos with the targeted users.\footnote{See \url{https://hivo.co/blog/creating-descriptive-titles-for-content-with-ai-a-how-to-guide}.} For instance, in a video featuring peaceful natural scenery—trees and flowing water—the AI-generated title, ``Enjoy the Beauty of Nature \#ScenicNature," captures the general theme but lacks specificity. In contrast, a human-revised AI-generated title,``Lush Mountains and Flowing Streams: Embrace Nature’s Serenity," offers greater lexical richness by adding more descriptors. Nouns like ``Mountains" and ``Streams" highlight the visual elements of the video content, while descriptive terms such as ``Lush" and ``Flowing" convey the element states, enhancing lexical density and variation. These context-specific enhancements allow the title to better describe and align with the video content, making it easier for the recommender system to interpret the video for more accurate content-user matching \citep{panniello2016research}. 

\begin{table}[htbp]\centering \footnotesize
\caption{Results of Lexical Richness Analysis (Matched Sample)}
\label{tab:lexical_richness_results}
\begin{threeparttable}
\begin{tabular}{@{\extracolsep{5pt}}lccc}
\\[-1.8ex]\hline 
\hline \\[-1.8ex] 
Dependent Variable & \textit{LexicalDensity} & \textit{LexicalVariation} & \textit{Entropy} \\
 & (1) & (2) & (3) \\
\hline \\[-1.8ex] 
\textit{Treat} & 0.016$^{***}$ & 0.017$^{***}$ & 0.011$^{***}$ \\
 & (0.0002) & (0.0002) & (0.0007) \\
\textit{Treat} * \textit{Similarity} & -0.245$^{***}$ & -0.301$^{***}$ & -0.442$^{***}$ \\
 & (0.0002) & (0.0002) & (0.0007) \\
\hline \\[-1.8ex]
Control Baseline (Mean) & 0.509 & 0.812 & 3.889 \\
Controls & YES & YES & YES \\
Observations & 3,885,089 & 3,885,089 & 3,885,089 \\
R-square & 0.621 & 0.674 & 0.803 \\
\hline 
\hline
\end{tabular} 
\begin{tablenotes}\scriptsize
\item \textit{Notes:} $^{***}$\textit{p}$<$0.01, $^{**}$\textit{p}$<$0.05, $^{*}$\textit{p}$<$0.1. Values in parentheses are robust standard errors.
\end{tablenotes}
\end{threeparttable}
\end{table}

To investigate why AI-generated titles led to high-quality but dissimilar titles, we surveyed 1,925 treatment group users in August. This survey featured open questions on the usage of AI-generated titles. The qualitative feedback highlighted an \textit{inspiration effect}, where users perceived AI-generated titles as a creative catalyst. For example, one user noted, “\textit{It’s already great; it may not always be precise, but it provides some inspiration for our posts!}” Another mentioned its help with “writer’s block,” saying, “\textit{When I can’t think of anything, it helps a bit.}” This evidence highlights the role of AI-generated titles as catalyst in content creation,\footnote{Similar arguments can be found in \url{https://www.mckinsey.com/capabilities/growth-marketing-and-sales/our-insights/how-generative-ai-can-boost-consumer-marketing}.} motivating producers to re-write or come up withtheir own titles. These findings align with recent studies on GAI and creativity \citep{art2024}.

\section{Additional Analyses and Robustness Tests}\label{sec:additional}

This section is devoted to further discussions and analyses to supplement our main results. The detailed regression results are relegated to Appendix \ref{appendixrobust}.

\subsubsection*{Viewership Diversity.} In the main text, we focus on the economic impact of AI-generated titles on content consumption, particularly consumption quantity (e.g., number of valid watch). However, their implications can be multifold, encompassing both quantity and diversity. This section analyzed how the access to AI-generated titles affects video viewership diversity using the Herfindahl-Hirschman Index (HHI), a widely used measure of market concentration in economic and antitrust analyses \citep{narayanan2009matter}. HHI is calculated by squaring the market share of each entity and summing the results, with values ranging from close to 0 to 1. A lower HHI indicates a more competitive environment, while a higher HHI signals dominance by one or a few large entities. In our analysis, we used valid watches to calculate HHI. The index ranged from 1/N to 1, where N was the total number of videos in our context. As shown in Table \ref{tab:platform_hhi_analysis}, the treatment group with access to AI-generated titles had a significantly lower HHI of 0.0002 compared to 0.0003 in the control group, representing a 50\% reduction in platform-level HHI. This indicates a substantial increase in viewership diversity, aligning with our earlier finding that AI-generated titles disproportionately benefited low-skilled producers. These results are consistent with recent GAI studies \citep{art2024}
and contribute to the growing body of research on GAI's impact on socioeconomic inequality \citep{capraro2024impact}.

\subsubsection*{Channel Analysis.}\label{sec:human_ai_channel}
In Section \ref{sec:Research Context}, we have discussed that organic and search-oriented recommendations are two main video recommendation channels on Platform A. Building on this, we next analyzed viewership outcomes for each channel separately and replicated the main analysis in Equation \eqref{formula:main_ols}. As shown in Table \ref{tab:search_oriented_recommendations} and Table \ref{tab:organic_recommendations}, the positive coefficients of $Treat$ are qualitatively aligned with the results in Table \ref{tab:ai_generated_titles}. These results indicate that AI-generated titles improve content consumption across both channels, reinforcing the effectiveness of AI-generated titles in enhancing user-video matching.

\subsubsection*{Impact on Content Production.} One potential explanation for the boosted viewership is that access to AI-generated titles changes users' video production behavior. For example, producers with access to AI-generated titles might spend more time refining each video’s content and producing fewer, but higher-quality, videos. To examine this possibility, we conducted a producer-level $t$-test comparing both the total number of videos and the average time gaps (in hours) between videos produced by each producer in the treatment and control groups during the treatment period. The results, shown in Table \ref{tab:video_production_comparison}, indicate no statistically significant difference in the number of videos produced and time gaps between videos between the two groups. These results suggest that the increase in viewership in the treatment group is unlikely to be driven by a change in producers' video production behavior.

\subsubsection*{Alternative Operationalization of Variables.}\label{sec:operationalization} We employed several alternative operationalizations of variables to ensure the robustness of our results. First, we used the number of watches ($Watch_{ij}$), complete watches\footnote{A complete watch is coded as 1 when the watch duration exactly matches the full video duration.} ($CompleteWatch_{ij}$), and the number of likes ($Like_{ij}$) for video $j$ of producer $i$ as alternative dependent variables for viewership outcomes. The ITT results presented in Table \ref{tab:itt_robust_alterdv} and \ref{tab:heter_itt_robust_alterdv}, and LATE results shown in Table \ref{tab:late_alternative_dv} and \ref{tab:heter_late_alternative_dv} of Online Appendix \ref{appendixalterdv}, are qualitatively consistent with our main findings.

Second, in Section \ref{data_and_variables}, $LowSkill_{i}$ is coded as 1 if producer $i$'s number of followers is below the median. To test robustness, we alternatively coded $LowSkill_{i}$ as 1 if the cumulative number of videos posted by producer $i$ exceeds the median (157). The ITT and LATE results, presented in Table \ref{tab:alternative_lowskill}, remain qualitatively consistent with our main findings.

Third, in Section \ref{data_and_variables}, $Adopt_{ij}$ is coded as 1 for exact matches between AI-generated titles and posted video titles. As a robustness test, we coded $Adopt_{ij}$ as 1 if the cosine similarity between the AI-generated title and the video's posted title met or exceeded specified thresholds (i.e., 95\%, 90\%, 85\%, and 80\%). The LATE results in Table \ref{tab:adopting_ai_titles_matching_outcomes}, using $ValidWatch_{ij}$ as the dependent variable, are qualitatively aligned with our findings. 

Fourth, we employed the Levenshtein algorithm as an alternative method to calculate the textual similarity ($Similarity_{ij}$) between video titles and AI-generated titles in Section \ref{sec:human-ai collaboration}. The Levenshtein algorithm, also known as the edit-distance metric, measures the minimum number of insertions, deletions, and substitutions required to transform one string into another, with each operation having a cost of one. Higher edit counts indicate lower similarity between AI-generated titles and actual video titles. The regression results for viewership outcomes and title lexical richness, presented in Tables \ref{tab:model_estimation_levenshtein} and \ref{tab:lexical_richness_levenshtein}, are qualitatively consistent with our findings.

\section{Conclusion and Discussion}\label{sec:conclu}
Previous research has shown that AI-generated content, such as advertisements, effectively engages users by enhancing content quality. However, the value of AI-generated content that does not directly interact with users, such as metadata, remains less understood. To address this gap, we conducted a randomized field experiment on a short-video platform where AI-generated titles, displayed in the bottom left corner, were rarely noticed by users. This setup allows us to isolate the effect of AI-generated metadata without direct user interaction. Our results show that AI-generated titles significantly boosted video consumption. Specifically, access to AI-generated video titles increased valid watches by 1.6\% and watch duration by 0.9\%. Further analysis suggests that this impact was primarily driven by addressing metadata scarcity, as evidenced by a notable increase in title availability. Moreover, we find that this effect was amplified for utilitarian videos and those produced by low-skilled creators, i.e., groups more affected by metadata sparsity. Specifically, utilitarian-content videos in the treatment group saw a relative decrease of 3.1\% in valid watches and 3.0\% in watch duration compared to hedonic-content videos. In contrast, low-skilled producers experienced an additional increase of 1.6\% in valid watches, and 1.3\% in watch duration. Mechanism analysis further indicates that AI-generated titles enhanced user-video matching accuracy. However, AI-generated titles were often of lower quality than human-generated ones, and only human-revised AI titles improved engagement further, highlighting the potential of combining AI with human input for better outcomes.

\subsection{Practical Implications}

Our results shed light on several important managerial implications. First, our study demonstrates that AI-generated metadata can significantly boost content discovery by improving user-content matching through mitigating metadata sparsity. Therefore, we encourage platform owners to invest in GAI tools that generate metadata, which can address operational challenges related to sparse metadata. While much of the focus has been on GAI’s ability to create user-facing content (e.g., advertisements and articles), our results emphasize the equally crucial role of AI-generated metadata in improving platform operations and boosting content discovery. This is relevant for platforms where content consumption is primarily driven by recommendations, such as UGC platforms and e-commerce sites.

Second, by demonstrating that AI-generated metadata disproportionately benefits low-skilled producers and hedonic-content videos, our work reveals the importance of tailoring platform strategies to support those most affected by metadata sparsity. Platforms should consider focusing their efforts on these segments when deciding whether to scale up the implementation of AI-generated metadata tools and how to maximize their effectiveness. For example, platforms can prioritize rolling out these tools to groups more affected by metadata sparsity, such as novice producers, to generate the most immediate and noticeable impact on content consumption.

Third, while GAI tools streamline video title generation, our results show that the quality of these AI-generated titles often falls short of human-generated ones. Therefore, rather than automatically integrating AI-generated titles into their recommender systems, platforms are encouraged to display these titles to content producers and provide the option to modify or enhance the metadata. Building on this, platforms are also recommended to incentivize content producers to revise or enhance AI-generated metadata rather than adopting them automatically. For example, platforms could place prominent reminders or offer traffic awards, to encourage producers to revise AI-generated titles. This can allow producers to inject their creativity and domain expertise to improve metadata. 
Additionally, our results show that improved titles with greater linguistic richness were associated with better viewership outcomes. To capitalize on this, platforms should consider offering workshops or tutorials to equip content producers with the skills needed to effectively use AI tools. By training producers to create more contextually rich and detailed metadata, platforms can enhance content discoverability and drive higher engagement.

\subsection{Limitations and Future Research}
The limitations of our work open up interesting avenues for future research. First, while we focus on the value of AI-generated metadata in improving user-content matching, we only explore content-based metadata. Future research could explore the role of user-related metadata, such as AI-generated user profiles. GAI can generate synthetic user profiles based on minimal inputs or demographic similarities, which may help the system to better predict user preferences and improve matching accuracy. Additionally, examining how different types of AI-generated metadata (e.g., content and user metadata) interact or complement each other also deserves exploration. Second, in our study, the GAI algorithm generated titles solely based on video content, without incorporating producer attributes or their historical content. Future research could explore how to design and improve AI-generated metadata to induce stronger effects. One potential direction is to incorporate producer-specific data, such as frequently used keywords from past videos or audience engagement patterns, to generate personalized AI-generated titles.

\bibliographystyle{ormsv080}
\bibliography{main}
\clearpage

%%%%%%%%%%%%%%%%%%%%%%%%%%%%%%%%%%%%%%%%%%%%%%%%%%%%%%%%%%%%%%%%%%%%%%%%%%%%%%%%%%
\begin{appendices}
\setcounter{page}{1}
\setcounter{footnote}{0}

\section{Pearson Correlation Matrix of Focal Variables}\label{appendix1}
\begin{table}[htbp]
\centering\footnotesize 
\caption{Pearson Correlation Matrix of Focal Variables}
\label{tab:correlation_matrix}
\resizebox{\textwidth}{!}{%
\begin{tabular}{@{\extracolsep{5pt}}lccccccccccccccc}
\hline
\hline
 & (1) & (2) & (3) & (4) & (5) & (6) & (7) & (8) & (9) & (10) & (11) & (12) & (13) & (14) \\
\hline
(1) \textit{Treat} & 1 &  &  &  &  &  &  &  &  &  &  &  &  &  \\
(2) \textit{Adopt} & 0.36 & 1 &  &  &  &  &  &  &  &  &  &  &  &  \\
(3) \textit{ValidWatch} & 0.00 & -0.01 & 1 &  &  &  &  &  &  &  &  &  &  &  \\
(4) \textit{WatchDuration} & 0.00 & -0.01 & 0.80 & 1 &  &  &  &  &  &  &  &  &  &  \\
(5) \textit{Utilitarian} & 0.01 & -0.01 & 0.01 & 0.01 & 1 &  &  &  &  &  &  &  &  &  \\
(6) \textit{LowSkill} & 0.00 & 0.02 & -0.02 & -0.02 & 0.00 & 1 &  &  &  &  &  &  &  &  \\
(7) \textit{Follower} & 0.00 & -0.02 & -0.17 & 0.13 & 0.01 & -0.06 & 1 &  &  &  &  &  &  &  \\
(8) \textit{KOL} & 0.00 & -0.02 & 0.02 & 0.01 & 0.01 & -0.07 & 0.12 & 1 &  &  &  &  &  &  \\
(9) \textit{Experience} & 0.00 & 0.09 & 0.00 & 0.00 & -0.05 & -0.29 & 0.04 & 0.06 & 1 &  &  &  &  &  \\
(10) \textit{Following} & 0.00 & 0.08 & 0.01 & -0.01 & 0.02 & -0.42 & -0.01 & 0.00 & 0.15 & 1 &  &  &  &  \\
(11) \textit{Multihome} & 0.00 & -0.03 & 0.01 & 0.01 & -0.01 & -0.02 & 0.02 & 0.03 & 0.11 & -0.05 & 1 &  &  &  \\
(12) \textit{Female} & 0.00 & 0.01 & -0.01 & -0.01 & -0.03 & -0.04 & -0.01 & -0.03 & 0.04 & -0.03 & 0.06 & 1 &  &  \\
(13) \textit{PublicVisible} & 0.00 & -0.04 & 0.01 & 0.01 & 0.03 & -0.06 & 0.01 & 0.02 & -0.03 & 0.06 & -0.01 & -0.09 & 1 &  \\
(14) \textit{VideoDuration} & 0.00 & 0.02 & 0.03 & 0.04 & 0.05 & -0.01 & 0.03 & 0.01 & 0.06 & 0.05 & 0.00 & -0.06 & 0.02 & 1 \\
\hline
\hline
\end{tabular}%
}
\end{table}

\section{Propensity Score Matching (PSM) Results for Titled Videos}\label{appendixpsm}
From the total of 10,738,984 videos, we first excluded 3,444,235 videos which do not have titles from both the treatment and control groups. Additionally, we removed 2,226,922 titled videos (51.56\%) in our treatment group that do not receive AI-generated titles due to algorithmic limitations. The filtered sample consisted of 2,092,219 videos in the treatment group and 2,975,608 in the control group. Next, we used the Propensity Score Matching (PSM) method to identify a sample that was similar in observed characteristics during the pre-treatment period. The matching followed a two-step procedure: first, we ran a logit regression using the pre-treatment variables (i.e., all the moderating and control variables mentioned in Section \ref{data_and_variables}) and obtained predicted propensity scores for each unit. Second, we employed a one-to-many radius matching algorithm where all control units for which the propensity scores fall within a pre-defined radius (also known as caliper) from the propensity scores of the treatment units are matched. This ensures multiple matches for each treated unit. We then applied weights in the subsequent analysis to account for this one-to-many structure. Next, we obtained a new sample after discarding unmatched units (3,885,089 videos are matched). To evaluate the matching quality, we performed $t$-tests of equality of means before and after matching to verify whether our matching has successfully balanced the attributes between the treatment and control
group videos. The results in Table \ref{tab:psm_radius} show that the mean differences between the groups were no longer statistically significant, indicating that the matching process successfully reduced bias associated with observable attributes.

\begin{table}[htbp]
    \centering 
    \footnotesize  
    \caption{Differences in Mean Before and After Matching (PSM, Radius Matching)} 
    \label{tab:psm_radius}
    \begin{threeparttable}
    \resizebox{\textwidth}{!}{%
    \begin{tabular}{@{\extracolsep{5pt}}lcrrrrr}
    \hline\hline
    Variable & Sample & Mean Treated & Mean Control & \%Bias & $T$-statistics & $P$-value \\
    \hline
    \textit{Utilitarian}    & Before-matched & 0.065 & 0.069 & -1.900  & -20.650  & 0.000 \\
                            & After-matched  & 0.065 & 0.073 & -3.400  & -0.050   & 0.960 \\
    \textit{LowSkill}       & Before-matched & 0.523 & 0.471 & 10.200  & 113.610  & 0.000 \\
                            & After-matched  & 0.523 & 0.528 & -1.000  & -0.010   & 0.989 \\
    Ln(\textit{Follower})   & Before-matched & 2.975 & 2.710 & 28.700  & 315.750  & 0.000 \\
                            & After-matched  & 2.975 & 2.905 & 3.800   & 0.110    & 0.911 \\
    \textit{KOL}            & Before-matched & 0.004 & 0.008 & -4.500  & -48.470  & 0.000 \\
                            & After-matched  & 0.004 & 0.005 & -1.100  & -0.020   & 0.986 \\
    \textit{Experience}     & Before-matched & 2.642 & 2.072 & 31.900  & 354.270  & 0.000 \\
                            & After-matched  & 2.642 & 2.505 & 7.700   & 0.110    & 0.915 \\
    Ln(\textit{Following})  & Before-matched & 2.907 & 2.989 & -8.200  & -89.620  & 0.000 \\
                            & After-matched  & 2.907 & 2.881 & 2.600   & 0.040    & 0.968 \\
    \textit{Multihome}      & Before-matched & 0.720 & 0.768 & -10.900 & -121.070 & 0.000 \\
                            & After-matched  & 0.720 & 0.735 & -3.400  & -0.050   & 0.962 \\
    \textit{Female}         & Before-matched & 0.658 & 0.658 & -0.100  & -1.500   & 0.000 \\
                            & After-matched  & 0.658 & 0.649 & 1.900   & 0.030    & 0.979 \\
    \textit{PublicVisible}  & Before-matched & 0.921 & 0.949 & -11.200 & -126.320 & 0.000 \\
                            & After-matched  & 0.921 & 0.930 & -3.400  & -0.040   & 0.965 \\
    Ln(\textit{VideoDuration}) & Before-matched & 0.327 & 0.269 & 16.000  & 177.980  & 0.000 \\
                               & After-matched  & 0.327 & 0.337 & -2.600  & -0.040   & 0.971 \\
    \textit{Cover}          & Before-matched & 0.135 & 0.196 & -16.300 & -178.890 & 0.000 \\
                            & After-matched  & 0.135 & 0.160 & -6.700  & -0.100   & 0.918 \\
    \textit{ContentType}    & Before-matched & 0.828 & 0.660 & 39.200  & 425.900  & 0.000 \\
                            & After-matched  & 0.828 & 0.850 & -5.100  & -0.080   & 0.934 \\
    \hline\hline
    \end{tabular}%
    }
    \begin{tablenotes}\scriptsize  
    \item Notes: ***$p<$0.01; **$p<$0.05; *$p<$0.1. \%Bias measures the standardized difference in covariate means, scaled by the standard deviation of the sample including both treatment and control groups. The closer it is to zero, the better the balance between the treatment and control groups. Values for \textit{Follower} and \textit{Following} have been scaled.
    \end{tablenotes} 
    \end{threeparttable}
\end{table}

\section{PSM Results for Videos with Low Textual Similarity}\label{appendixpsm2}
From the 2,092,219 titled videos in the treatment group and 2,975,608 in the control group (as described in Online Appendix \ref{appendixpsm}), we first excluded 1,532,819 treatment videos with cosine similarity to AI-generated titles higher than 20\%. This filtering resulted in 559,400 videos in the treatment group and 2,975,608 in the control group. Using the Propensity Score Matching (PSM) method with a radius matching algorithm, we identified a matched sample based on pre-treatment characteristics (i.e., all moderating and control variables in Section \ref{data_and_variables}), following the detailed procedure in Online Appendix \ref{appendixpsm}. After discarding unmatched units, the final matched sample included 3,533,720 videos. Post-matching $t$-tests of equality of means, shown in Table \ref{tab:psm_radius2}, indicate that the mean differences between the treatment and control groups were no longer statistically significant, confirming that the matching process effectively reduced bias.

\begin{table}[H]
    \centering 
    \footnotesize  
    \caption{Differences in Mean Before and After Matching (PSM, Radius Matching)} 
    \label{tab:psm_radius2}
    \begin{threeparttable}
    \resizebox{\textwidth}{!}{%
    \begin{tabular}{@{\extracolsep{5pt}}lcrrrrr}
    \hline\hline
    Variable & Sample & Mean Treated & Mean Control & \%Bias & $T$-statistics & $P$-value \\
    \hline
    \textit{Utilitarian}    & Before-matched & 0.065 & 0.069 & -1.900  & -20.650  & 0.000 \\
                            & After-matched  & 0.066 & 0.069 & -1.100  & -0.040   & 0.966 \\
    \textit{LowSkill}       & Before-matched & 0.523 & 0.471 & 10.200  & 113.610  & 0.000 \\
                            & After-matched  & 0.533 & 0.485 & 9.700   & 0.360    & 0.717 \\
    Ln(\textit{Follower})   & Before-matched & 2.975 & 2.710 & 28.700  & 315.750  & 0.000 \\
                            & After-matched  & 2.716 & 2.700 & 1.700   & 0.070    & 0.948 \\
    \textit{KOL}            & Before-matched & 0.004 & 0.008 & -4.500  & -48.470  & 0.000 \\
                            & After-matched  & 0.006 & 0.007 & -1.400  & -0.060   & 0.955 \\
    \textit{Experience}     & Before-matched & 2.642 & 2.072 & 31.900  & 354.270  & 0.000 \\
                            & After-matched  & 2.541 & 2.075 & 25.600  & 0.930    & 0.353 \\
    Ln(\textit{Following})  & Before-matched & 2.907 & 2.989 & -8.200  & -89.620  & 0.000 \\
                            & After-matched  & 2.911 & 2.947 & -3.600  & -0.140   & 0.886 \\
    \textit{Multihome}      & Before-matched & 0.720 & 0.768 & -10.900 & -121.070 & 0.000 \\
                            & After-matched  & 0.777 & 0.767 & 2.600   & 0.100    & 0.922 \\
    \textit{Female}         & Before-matched & 0.658 & 0.658 & -0.100  & -1.500   & 0.000 \\
                            & After-matched  & 0.660 & 0.657 & 0.700   & 0.030    & 0.980 \\
    \textit{PublicVisible}  & Before-matched & 0.921 & 0.949 & -11.200 & -126.320 & 0.000 \\
                            & After-matched  & 0.959 & 0.952 & 3.200   & 0.130    & 0.898 \\
    Ln(\textit{VideoDuration}) & Before-matched & 0.327 & 0.269 & 16.000  & 177.980  & 0.000 \\
                               & After-matched  & 0.277 & 0.263 & 4.000   & 0.150    & 0.881 \\
    \textit{Cover}          & Before-matched & 0.135 & 0.196 & -16.300 & -178.890 & 0.000 \\
                            & After-matched  & 0.226 & 0.194 & 7.800   & 0.280    & 0.777 \\
    \textit{ContentType}    & Before-matched & 0.828 & 0.660 & 39.200  & 425.900  & 0.000 \\
                            & After-matched  & 0.648 & 0.647 & 0.200   & 0.010    & 0.994 \\
    \hline\hline
    \end{tabular}%
    }
    \begin{tablenotes}\scriptsize  
    \item Notes: ***$p<$0.01; **$p<$0.05; *$p<$0.1. \%Bias measures the standardized difference in covariate means, scaled by the standard deviation of the sample including both treatment and control groups. The closer it is to zero, the better the balance between the treatment and control groups. Values for \textit{Follower} and \textit{Following} have been scaled.
    \end{tablenotes} 
    \end{threeparttable}
\end{table}

\section{Additional Analyses and Robustness Tests}\label{appendixrobust}
\vspace{1em}
\subsection{Results for Viewership Diversity Analysis}\label{appendixdiversity}
For analyses reported in the main text (as explained in Section \ref{sec:direct_effect}), we focused on the influence of AI-generated titles on video viewership. Here, we extend this analysis to explore their impact on viewership diversity. We first divided the total videos in our sample into two groups: videos produced by treatment group users and videos by control group users. We then calculated HHI based on valid watches to measure viewership concentration for each group, defined as:
\begin{equation}
\text{HHI} = \sum_{j=1}^{N} s_j^2 \label{eq:HHI1}
\end{equation}
where \( s_j \) represents the market share of video \( j \) within its respective group, calculated as:
\begin{equation}
s_i = \frac{V_j}{\sum_{j=1}^{N} V_j} \label{eq:HHI2}
\end{equation}
where \( V_j \) is the number of valid watches for video \( j \), and the denominator is the total number of valid watches across all \( N \) videos within the group.

We used 1000 bootstrap iterations for each group to calculate HHI. In each iteration, we randomly sampled videos with replacements from each group to create a bootstrap sample and calculated the HHI for this sample using equations \eqref{eq:HHI1} and \eqref{eq:HHI2}. This process yielded 1000 HHI values for both the treatment and control groups, enabling a bootstrap-based significance test to assess whether the difference in viewership concentration between the two groups was statistically significant. The results in Table \ref{tab:platform_hhi_analysis} show that the treatment group with access to AI-generated titles had a significantly lower HHI of 0.0002 compared to 0.0003 in the control group ($p$ $< 0.05$), representing a 50\% reduction in platform-level HHI. A lower HHI indicates a more evenly distributed viewership across videos, reflecting greater diversity. This substantial increase in viewership diversity is consistent with our earlier finding in Section \ref{sec:direct_effect} that AI-generated titles disproportionately benefited lower-skilled producers.

\begin{table}[htbp]
\centering
\footnotesize
\caption{Results of Platform-level HHI Analysis}
\label{tab:platform_hhi_analysis}
\begin{threeparttable}
\begin{tabular}{lcccc}
\hline
 & Treatment Group & Control Group & Difference & $P$-value \\
\hline
 & (1) & (2) & (3) & (4) \\
 \hline
 & 0.0003 & 0.0002 & -0.0001 & $<0.05$ \\
\hline
\end{tabular}
\begin{tablenotes}\scriptsize
\item \textit{Notes:} We perform 1000 bootstrap iterations to compute the HHI and p-value.
\end{tablenotes}
\end{threeparttable}
\end{table}

\subsection{Results for Channel Analysis}\label{appendixchannel}
In Section \ref{sec:mechanism}, we suggested that the positive effect of AI-generated titles on video viewership is attributed to enhanced recommendation accuracy. Theoretically, if true, this effect should benefit both recommendation channels on Platform A (organic and search-oriented recommendations). To test this, we used detailed video consumption data to identify the recommendation source for each watch. Next, we aggregated the valid watch counts for each video by channel and replicated the analysis in Equation \eqref{formula:main_ols} for each channel. The results, presented in Table \ref{tab:search_oriented_recommendations} and \ref{tab:organic_recommendations}, show a 0.3\% increase in both valid watches and watch duration for treatment group videos in the search-oriented channel, and a stronger 1.5\% increase in valid watches with a 0.9\% increase in watch duration in the organic channel. These findings validate the effectiveness of AI-generated titles in enhancing viewership metrics across both channels.

\begin{table}[H]\centering \footnotesize  
\caption{Results for Search-Oriented Recommendations} 
\label{tab:search_oriented_recommendations} 
\begin{threeparttable}
\begin{tabular}{@{\extracolsep{5pt}}lcc} 
\\[-1.8ex]\hline 
\hline \\[-1.8ex] 
Dependent Variable & \textit{ValidWatch} & \textit{WatchDuration} \\
\\[-1.8ex] & (1) & (2) \\
\hline \\[-1.8ex] 
\textit{Treat} & 0.003$^{***}$ & 0.003$^{***}$ \\ 
 & (0.0003) & (0.0002) \\ 
\hline
Controls & YES & YES \\ 
Observations & 10,738,984 & 10,738,984 \\ 
R-square & 0.166 & 0.165 \\ 
\hline 
\hline \\[-1.8ex] 
\end{tabular}
\begin{tablenotes}\scriptsize
\item \textit{Notes:} $^{***}$\textit{p}$<$0.01; $^{**}$\textit{p}$<$0.05; $^{*}$\textit{p}$<$0.1. Values in parentheses are robust standard errors.
\end{tablenotes}
\end{threeparttable}
\end{table}

\begin{table}[H]\centering \footnotesize  
\caption{Results for Organic Recommendations} 
\label{tab:organic_recommendations} 
\begin{threeparttable}
\begin{tabular}{@{\extracolsep{5pt}}lcc} 
\\[-1.8ex]\hline 
\hline \\[-1.8ex] 
Dependent Variable & \textit{ValidWatch} &\textit{WatchDuration} \\
\\[-1.8ex] & (1) & (2) \\
\hline \\[-1.8ex] 
\textit{Treat} & 0.015$^{***}$ & 0.009$^{***}$ \\ 
 & (0.001)  & (0.001)  \\ 
\hline
Controls & YES & YES \\ 
Observations & 10,738,984 & 10,738,984 \\ 
R-square & 0.295 & 0.321 \\ 
\hline 
\hline \\[-1.8ex] 
\end{tabular}
\begin{tablenotes}\scriptsize
\item \textit{Notes:} $^{***}$\textit{p}$<$0.01; $^{**}$\textit{p}$<$0.05; $^{*}$\textit{p}$<$0.1. Values in parentheses are robust standard errors.
\end{tablenotes}
\end{threeparttable}
\end{table}

\subsection{Results for Video Production Comparison }\label{appendixalterproduction}
In Section \ref{sec:mechanism}, we suggested that the positive effect of AI-generated titles on video viewership is attributed to enhanced recommendation accuracy. An alternative explanation for the increased viewership could be a supply-side shift in production behavior. Specifically, treatment producers with access to AI-generated titles might have altered their production behavior by focusing on producing fewer but potentially higher-quality videos. This shift could theoretically enhance perceived video quality, thereby increasing viewers' content consumption. To examine this possibility, we conducted a producer-level $t$-test comparing the total number of videos and average time gaps (in hours) between videos produced by each producer in the treatment and control groups during the treatment period. As shown in Table \ref{tab:video_production_comparison}, there is no statistically significant difference in the number of videos produced (\textit{p}-value $>$ 0.1) and the time gaps between videos (\textit{p}-value $>$ 0.1) across the two groups. These results suggest that the increase in viewership in the treatment group is unlikely to be driven by a change in producers' video production behavior.

\begin{table}[htbp]
\centering
\footnotesize
\caption{Comparison of Video Production Between Treatment and Control Producers}
\label{tab:video_production_comparison}
\begin{tabular}{lccccc}
\hline
Variable & Treatment Group & Control Group & Difference & \textit{P}-value \\
\hline
\textit{NumVideo}          & 5.246                   & 5.240                  & 0.006                & 0.660                     \\
\textit{TimeGap}   & 51.241                  & 51.249                 & -0.008               & 0.928                     \\
\hline
\end{tabular}
\end{table}

\subsection{Results for Alternative Operationalization of Variables}\label{appendixalterdv}
To ensure the robustness of our main findings, we employed alternative operationalizations for key variables in our analysis. Given that the observed positive effect of AI-generated titles may vary depending on how viewership and adoption are defined, testing these alternative definitions allows us to verify the consistency and generalizability of our results. First, for viewership, we used the number of watches($Watch_{ij}$), complete watches ($CompleteWatch_{ij}$), and the number of likes ($Like_{ij}$) per video as alternative dependent variables. These measures capture different aspects of viewer engagement and may offer additional insights into how AI-generated titles impact a range of interactive forms of engagement. The ITT and LATE results, shown in Tables \ref{tab:itt_robust_alterdv}-\ref{tab:heter_late_alternative_dv}, are qualitatively consistent with our findings in Section \ref{sec:intention_to_treat} and \ref{sec:LATE}, confirming that AI-generated titles positively impact users' content consumption and engagement.

\begin{table}[htbp]\centering \footnotesize  
\caption{ITT Results Using Alternative Dependent Variables} 
\label{tab:itt_robust_alterdv} 
\begin{threeparttable}
\begin{tabular}{@{\extracolsep{5pt}}lccc} 
\\[-1.8ex]\hline 
\hline \\[-1.8ex] 
Dependent Variable & \textit{Watch} & \textit{CompleteWatch} & \textit{Like} \\
\\[-1.8ex] & (1) & (2) & (3) \\
\hline \\[-1.8ex] 
\textit{Treat} & 0.015$^{***}$ & 0.007$^{***}$ & 0.021$^{***}$ \\ 
 & (0.001) & (0.001) & (0.001) \\ 
\hline 
Controls & YES & YES & YES \\ 
Observations & 10,738,984 & 10,738,984 & 10,738,984 \\ 
R-square & 0.313 & 0.252 & 0.297 \\ 
\hline 
\hline \\[-1.8ex] 
\end{tabular}
\begin{tablenotes}\scriptsize
\item \textit{Notes:} $^{***}$\textit{p}$<$0.01; $^{**}$\textit{p}$<$0.05; $^{*}$\textit{p}$<$0.1. Values in parentheses are robust standard errors.
\end{tablenotes}
\end{threeparttable}
\end{table}

\begin{table}[htbp]\centering \footnotesize  
\caption{Heterogeneous ITT Results Using Alternative Dependent Variables} 
\label{tab:heter_itt_robust_alterdv} 
\begin{threeparttable}
\begin{tabular}{@{\extracolsep{5pt}}lcccccc} 
\\[-1.8ex]\hline 
\hline \\[-1.8ex] 
 & \multicolumn{2}{c}{\textit{Watch}} & \multicolumn{2}{c}{\textit{CompleteWatch}} & \multicolumn{2}{c}{\textit{Like}} \\
\cline{2-3} \cline{4-5} \cline{6-7} 
\\[-1.8ex] Dependent Variable & (1) & (2) & (3) & (4) & (5) & (6) \\
\hline \\[-1.8ex] 
\textit{Treat} & 0.017$^{***}$ & -0.0003 & 0.008$^{***}$ & -0.001 & 0.025$^{***}$ & 0.011$^{***}$ \\ 
 & (0.001) & (0.002) & (0.001) & (0.001) & (0.001) & (0.001) \\ 

\textit{LowSkill} &  & -0.911$^{***}$ &  & -0.623$^{***}$ &  & -0.849$^{***}$ \\ 
 &  & (0.002) &  & (0.001) &  & (0.001) \\ 
 
\textit{Treat} * \textit{Utilitarian} & -0.035$^{***}$ &  & -0.013$^{***}$ &  & -0.056$^{***}$ &  \\ 
 & (0.004) &  & (0.004) &  & (0.003) &  \\ 
\textit{Treat} * \textit{LowSkill} &  & 0.022$^{***}$ &  & 0.010$^{***}$ &  & 0.014$^{***}$ \\ 
 &  & (0.002) &  & (0.002) &  & (0.002) \\ 
\hline 
Controls & YES & YES & YES & YES & YES & YES \\ 
Observations & 10,738,984 & 10,738,984 & 10,738,984 & 10,738,984 & 10,738,984 & 10,738,984 \\ 
R-square & 0.313 & 0.252 & 0.252 & 0.190 & 0.297 & 0.258 \\ 
\hline 
\hline \\[-1.8ex] 
\end{tabular}
\begin{tablenotes}\scriptsize
\item \textit{Notes:} $^{***}$\textit{p}$<$0.01; $^{**}$\textit{p}$<$0.05; $^{*}$\textit{p}$<$0.1. Values in parentheses are robust standard errors. \textit{Utilitarian} is absorbed by the video category dummies in \textit{Controls} and therefore not reported in the table.
\end{tablenotes}
\end{threeparttable}
\end{table}

\begin{table}[htbp]\centering \footnotesize  
\caption{LATE Results for Main Effects Using Alternative Dependent Variables} 
\label{tab:late_alternative_dv} 
\begin{threeparttable}
\begin{tabular}{@{\extracolsep{5pt}}lccc} 
\\[-1.8ex]\hline 
\hline \\[-1.8ex] 
Dependent Variable & \textit{Watch} & \textit{CompleteWatch} & \textit{Like} \\
\\[-1.8ex] & (1) & (2) & (3) \\
\hline \\[-1.8ex] 
\textit{Adopt} & 0.063$^{***}$ & 0.029$^{***}$ & 0.091$^{***}$ \\ 
 & (0.004) & (0.004) & (0.004) \\ 
\hline 
Controls & YES & YES & YES \\ 
Observations & 10,738,984 & 10,738,984 & 10,738,984 \\ 
R-square & 0.313 & 0.252 & 0.297 \\ 
\hline 
\hline \\[-1.8ex] 
\end{tabular}
\begin{tablenotes}\scriptsize
\item \textit{Notes:} $^{***}$\textit{p}$<$0.01; $^{**}$\textit{p}$<$0.05; $^{*}$\textit{p}$<$0.1. Values in parentheses are robust standard errors.
\end{tablenotes}
\end{threeparttable}
\end{table}

\begin{table}[htbp]\centering \footnotesize  
\caption{Heterogeneous LATE Results Using Alternative Dependent Variables} 
\label{tab:heter_late_alternative_dv} 
\begin{threeparttable}
\begin{tabular}{@{\extracolsep{5pt}}lcccccc} 
\\[-1.8ex]\hline 
\hline \\[-1.8ex] 
 {Dependent Variable} & \multicolumn{2}{c}{\textit{Watch}} & \multicolumn{2}{c}{\textit{CompleteWatch}} & \multicolumn{2}{c}{\textit{Like}} \\
\cline{2-3} \cline{4-5} \cline{6-7} 
\\[-1.8ex]  & (1) & (2) & (3) & (4) & (5) & (6) \\
\hline \\[-1.8ex] 
\textit{Adopt} & 0.073$^{***}$ & -0.0008 & 0.033$^{***}$ & -0.006 & 0.107$^{***}$ & 0.051$^{***}$ \\ 
 & (0.005) & (0.007) & (0.004) & (0.006) & (0.004) & (0.006) \\ 

\textit{LowSkill} &  & -0.913$^{***}$ &  & -0.624$^{***}$ &  & -0.852$^{***}$ \\ 
 &  & (0.002) &  & (0.001) &  & (0.001) \\ 
 
\textit{Adopt} * \textit{Utilitarian} & -0.162$^{***}$ &  & -0.250$^{***}$ &  & -0.058$^{***}$ &  \\ 
 & (0.020) &  & (0.016) &  & (0.018) &  \\ 
\textit{Adopt} * \textit{LowSkill} &  & 0.089$^{***}$ &  & 0.052$^{***}$ &  & 0.042$^{***}$ \\ 
 &  & (0.008) &  & (0.007) &  & (0.008) \\ 
\hline 
Controls & YES & YES & YES & YES & YES & YES \\ 
Observations & 10,738,984 & 10,738,984 & 10,738,984 & 10,738,984 & 10,738,984 & 10,738,984 \\ 
R-square & 0.313 & 0.252 & 0.252 & 0.190 & 0.297 & 0.258 \\ 
\hline 
\hline \\[-1.8ex] 
\end{tabular}
\begin{tablenotes}\scriptsize
\item \textit{Notes:} $^{***}$\textit{p}$<$0.01; $^{**}$\textit{p}$<$0.05; $^{*}$\textit{p}$<$0.1. Values in parentheses are robust standard errors. \textit{Utilitarian} is absorbed by the video category dummies in \textit{Controls} and therefore not reported in the table.
\end{tablenotes}
\end{threeparttable}
\end{table}

Second, in Section \ref{data_and_variables}, $LowSkill_{i}$ is coded as 1 if producer $i$'s number of followers is below the median. For robustness checks, $LowSkill_{i}$ is coded as 1 if the cumulative number of videos posted by producer $i$ exceeded the median (313). The ITT and LATE results in Table \ref{tab:alternative_lowskill} are qualitatively consistent with our main findings.

\begin{table}[htbp]\centering \footnotesize  
\caption{Model Estimation Results Using Alternative Measures for Low-skilled Producers} 
\label{tab:alternative_lowskill} 
\begin{threeparttable}
\begin{tabular}{@{\extracolsep{5pt}}lcccc} 
\\[-1.8ex]\hline 
\hline \\[-1.8ex] 
 & \multicolumn{2}{c}{{ITT}} & \multicolumn{2}{c}{LATE} \\
\cline{2-3} \cline{4-5} 
\\[-1.8ex] Dependent Variable & $ValidWatch$ & $WatchDuration$ & $ValidWatch$ & $WatchDuration$ \\
 & (1) & (2) & (3) & (4) \\
\hline \\[-1.8ex] 
\textit{Treat} & 0.006$^{***}$ & -0.0002 & 0.025$^{***}$ & -0.0005 \\ 
 & (0.001) & (0.001) & (0.006) & (0.005) \\ 

\textit{LowSkill} & -0.162$^{***}$ & -0.188$^{***}$ & -0.164$^{***}$ & -0.189$^{***}$ \\ 
 & (0.002) & (0.001) & (0.002) & (0.001) \\ 
 
\textit{Treat} * \textit{LowSkill} & 0.009$^{***}$ & 0.009$^{***}$ & 0.050$^{***}$ & 0.043$^{***}$ \\ 
 & (0.002) & (0.002) & (0.000) & (0.008) \\ 
\hline 
Controls & YES & YES & YES & YES \\ 
Observations & 10,738,984 & 10,738,984 & 10,738,984 & 10,738,984 \\ 
R-square & 0.204 & 0.223 & 0.203 & 0.222 \\ 
\hline 
\hline \\[-1.8ex] 
\end{tabular}
\begin{tablenotes}\scriptsize
\item \textit{Notes:} $^{***}$\textit{p}$<$0.01; $^{**}$\textit{p}$<$0.05; $^{*}$\textit{p}$<$0.1. Values in parentheses are robust standard errors.
\end{tablenotes}
\end{threeparttable}
\end{table}

Third, in Section \ref{sec:LATE}, \textit{Adopt} was defined as an exact match between the AI-generated title and the posted video title. However, exact matching may overlook cases where titles are similar but not identical. To address this, we coded \textit{Adopt} as 1 if the cosine similarity between the AI-generated title and the posted title met or exceeded thresholds of 95\%, 90\%, 85\%, and 80\%. This approach allows us to capture adoption behaviors more flexibly and examine whether even partial adoption drives viewership outcomes. Next, we re-estimated Equation \eqref{2sls_first} and \eqref{2sls_second} in our LATE analysis, using \textit{ValidWatch} as the dependent variable. The positive coefficients of \textit{Adopt} in Table \ref{tab:adopting_ai_titles_matching_outcomes} validate the effectiveness of AI-generated titles in enhancing content consumption outcomes.

\begin{table}[htbp]\centering \footnotesize  
\caption{Alternative Adoption Thresholds of AI-generated Titles and Effect on Content Consumption, DV = \textit{ValidWatch}} 
\label{tab:adopting_ai_titles_matching_outcomes} 
\begin{threeparttable}
\begin{tabular}{@{\extracolsep{5pt}}lcccc} 
\\[-1.8ex]\hline 
\hline \\[-1.8ex] 
Similarity Threshold & $>95\%$ & $>90\%$ & $>85\%$ & $>80\%$ \\
 & (1) & (2) & (3) & (4) \\
\hline \\[-1.8ex] 
\textit{Adopt} & 0.068$^{***}$ & 0.068$^{***}$ & 0.067$^{***}$ & 0.066$^{***}$ \\ 
 & (0.004) & (0.004) & (0.004) & (0.004) \\ 
Controls & YES & YES & YES & YES \\ 
Observations & 10,738,984 & 10,738,984 & 10,738,984 & 10,738,984 \\ 
R-square & 0.295 & 0.295 & 0.295 & 0.295 \\ 
\hline 
\hline \\[-1.8ex] 
\end{tabular}
\begin{tablenotes}\scriptsize
\item \textit{Notes:} $^{***}$\textit{p}$<$0.01; $^{**}$\textit{p}$<$0.05; $^{*}$\textit{p}$<$0.1. Values in parentheses are robust standard errors.
\end{tablenotes}
\end{threeparttable}
\end{table}

Lastly, we used the Levenshtein distance, also known as the edit distance, to compute the similarity between AI-generated and posted titles as an alternative to cosine similarity used in Section \ref{sec:human-ai collaboration}.  This metric calculates the minimum edits (e.g., insertions, deletions, substitutions) needed to transform one title into another, with higher distances indicating lower similarity. Using this alternative measure, we then replicated the analysis in Section \ref{sec:human-ai collaboration}. The positive coefficients of \textit{Treat} and negative coefficients of the interaction term between \textit{Treat} and \textit{Similarity}, as shown in Tables \ref{tab:model_estimation_levenshtein} and \ref{tab:lexical_richness_levenshtein}, are qualitatively aligned with our main analysis. Together, these alternative variable operationalizations indicate that our analyses are robust across different measures of consumption, adoption, and title similarity.

\begin{table}[H]\centering \footnotesize  
\caption{Model Estimation Results Using Levenshtein Algorithm (Matched Sample)} 
\label{tab:model_estimation_levenshtein} 
\begin{threeparttable}
\begin{tabular}{@{\extracolsep{5pt}}lcc} 
\\[-1.8ex]\hline 
\hline \\[-1.8ex] 
Dependent Variable & \textit{ValidWatch} & \textit{WatchDuration} \\
 & (1) & (2) \\
\hline \\[-1.8ex] 
\textit{Treat} & 0.251$^{***}$ & 0.208$^{***}$ \\ 
 & (0.002) & (0.002) \\ 
\textit{Treat} * \textit{Similarity} & -1.062$^{***}$ & -0.880$^{***}$ \\ 
 & (0.003) & (0.002) \\ 
\hline 
Controls & YES & YES \\ 
Observations & 3,885,089 & 3,885,089 \\ 
R-square & 0.385 & 0.396 \\ 
\hline 
\hline \\[-1.8ex] 
\end{tabular}
\begin{tablenotes}\scriptsize
\item \textit{Notes:} $^{***}$\textit{p}$<$0.01; $^{**}$\textit{p}$<$0.05; $^{*}$\textit{p}$<$0.1. Values in parentheses are robust standard errors.
\end{tablenotes}
\end{threeparttable}
\end{table}

\begin{table}[H]\centering \footnotesize  
\caption{Results of Lexical Richness Analysis Using Levenshtein Algorithm (Matched Sample)} 
\label{tab:lexical_richness_levenshtein} 
\begin{threeparttable}
\begin{tabular}{@{\extracolsep{5pt}}lccc} 
\\[-1.8ex]\hline 
\hline \\[-1.8ex] 
Dependent Variable & \textit{LexicalDensity} & \textit{LexicalVariation} & \textit{Entropy} \\
 & (1) & (2) & (3) \\
\hline \\[-1.8ex] 
\textit{Treat} & 0.031$^{***}$ & 0.034$^{***}$ & 0.030$^{***}$ \\ 
 & (0.0002) & (0.0002) & (0.001) \\ 
\textit{Treat} * \textit{Similarity} & -0.263$^{***}$ & -0.322$^{***}$ & -0.464$^{***}$ \\ 
 & (0.0003) & (0.0002) & (0.001) \\ 
\hline 
Controls & YES & YES & YES \\ 
Observations & 3,885,089 & 3,885,089 & 3,885,089 \\ 
R-square & 0.634 & 0.689 & 0.802 \\ 
\hline 
\hline \\[-1.8ex] 
\end{tabular}
\begin{tablenotes}\scriptsize
\item \textit{Notes:} $^{***}$\textit{p}$<$0.01; $^{**}$\textit{p}$<$0.05; $^{*}$\textit{p}$<$0.1. Values in parentheses are robust standard errors.
\end{tablenotes}
\end{threeparttable}
\end{table}

\end{appendices}
\end{document}